# LARO: Learned Acquisition and Reconstruction Optimization to accelerate Quantitative Susceptibility Mapping

Jinwei Zhang[1,2], Pascal Spincemaille[2], Hang Zhang[2,3], Thanh D. Nguyen[2], Chao Li[2,4], Jiahao Li[1,2], Ilhami Kovanlikaya[2], Mert R. Sabuncu[2,3], Yi Wang[1,2*]

[1]Department of Biomedical Engineering, Cornell University, Ithaca, New York

[2]Department of Radiology, Weill Medical College of Cornell University, New York, New York

[3]Department of Electrical and Computer Engineering, Cornell University, Ithaca, New York

[4]Department of Applied Physics, Cornell University, Ithaca, NY, USA

* Correspondence to:

Yi Wang, PhD.

Radiology, Weill Cornell Medicine

407 E 61st St, New York, NY 10065, USA.

E-mail: yiwang@med.cornell.edu

**ABSTRACT**

Quantitative susceptibility mapping (QSM) involves acquisition and reconstruction of a series of images at multi-echo time points to estimate tissue field, which prolongs scan time and requires specific reconstruction technique. In this paper, we present our new framework, called Learned Acquisition and Reconstruction Optimization (LARO), which aims to accelerate the multi-echo gradient echo (mGRE) pulse sequence for QSM. Our approach involves optimizing a Cartesian multi-echo k-space sampling pattern with a deep reconstruction network. Next, this optimized sampling pattern was implemented in an mGRE sequence using Cartesian fan-beam k-space segmenting and ordering for prospective scans. Furthermore, we propose to insert a recurrent temporal feature fusion module into the reconstruction network to capture signal redundancies along echo time. Our ablation studies show that both the optimized sampling pattern and proposed reconstruction strategy help improve the quality of the multi-echo image reconstructions. Generalization experiments show that LARO is robust on the test data with new pathologies and different sequence parameters. Our code is available at https://github.com/Jinwei1209/LARO.git.

**INTRODUCTION**

Quantitative magnetic resonance imaging (MRI) provides biomarkers for clinical assessment of diverse diseases, including T1 and T2 relaxation time (1,2), fat fraction (3), quantitative susceptibility mapping (QSM) (4), etc. For QSM, a multi-echo gradient echo (mGRE) pulse sequence is used to acquire signals at different echo times. A tissue-induced local magnetic field map can be obtained by fitting the acquired complex multi-echo signals (5,6). Then, a tissue susceptibility map can be computed using an inverse problem solver, such as regularized dipole inversion (7).

For QSM, the range of echo times needs to be large enough to cover both small and large susceptibilities in tissue (8), such as in the application of QSM in multiple sclerosis (MS), where QSM has been shown to be sensitive to myelin content as well as iron (8), both of which are modified in MS. However, limited scan time in clinics only allows for mGRE with a compromised spatial resolution, making visualization of smaller MS lesion more challenging. Overcoming this compromise is a major motivation for this work.

The significantly increased scan time of mGRE sequence can be partly overcome using classical acceleration techniques such as Parallel imaging (PI) (9,10), compressed sensing (CS) (11), or their combination (PI-CS) (12,13). Recently, deep learning has been used to optimize k-space sampling patterns from training data, such as in LOUPE (14) and its extension LOUPE-ST (15), experimental design with the constrained Cramer-Rao bound (OEDIPUS) (16) and greedy pattern selection (17). Building on these prior works, we propose here to learn an optimal sampling pattern to accelerate QSM acquisition and improve reconstruction quality.

Reconstruction from under-sampled measurements can be solved using regularization to exploit signal redundancies, such as low-rank and/or sparsity constraints (18-20). More recently, convolutional neural networks have been proposed for compressed sensing reconstruction. One popular neural network technique involves implementing the unrolled iterations of an optimization process, coupled with a learned regularizer, as in MoDL (21) and VarNet (21). These architectural designs have been applied to single-echo image reconstruction, and extended to dynamic image sequence reconstruction via cascaded (22) and recurrent networks (23). Recently QSM acquisition was accelerated using 2D incoherent Cartesian under-sampling and deep neural network reconstruction with a variable density sampling pattern manually designed and fixed across echoes (24).

We propose Learned Acquisition and Reconstruction Optimization (LARO) to further optimize the sampling pattern across echoes by inferring the temporal variation through adding a temporal dimension to LOUPE-ST (15) for the multi-echo case. Images are reconstructed accordingly using an unrolled reconstruction network based on alternating direction method of multipliers (ADMM) (25) to capture the signal evolution and compensate the aliasing patterns of mGRE images with a temporal feature fusion module.

In this work, the accelerated acquisition is not used to increase the spatial resolution but to instead accelerate the clinical protocol. This allows a retrospective analysis of the performance of the proposed method in existing clinical data. The resulting mGRE sampling pattern was also implemented in an mGRE sequence to prospectively acquire and reconstruct under-sampled k-space data for QSM. This work is extended from our conference paper (26) where preliminary retrospective results were shown as a proof of concept of LARO.

## THEORY

In QSM data acquisition, multi-echo k-space sampling with multiple receiver coils is modeled as:

$$b_{jk} = U_j F E_k s_j + n_{jk}, \qquad (1)$$

where $b_{jk}$ is the measured k-space data of the $k$-th receiver coil at the $j$-th echo time, with $N_C$ receiver coils and $N_T$ echo times, $U_j$ is the k-space under-sampling pattern at the $j$-th echo time, $F$ is the Fourier transform, $E_k$ is the sensitivity map of the $k$-th coil, $s_j$ is the complex image of the $j$-th coil to be reconstructed, and $n_{jk}$ is the acquisition noise, assumed to be Gaussian.

Having acquired $b_{jk}$ with fixed $U_j$, we aim at reconstructing all $s_j$ simultaneously with a cross-echo regularization loss $R(\{s_j\})$. Based on Eq. 1, a solution $\{\hat{s}_j\}$ can be obtained by solving the following optimization

$$\{\hat{s}_j\} = \underset{\{s_j\}}{\operatorname{argmin}} \, \mathrm{E}(\{s_j\}) = \underset{\{s_j\}}{\operatorname{argmin}} \sum_{j=1}^{N_T} \sum_{k=1}^{N_C} \|U_j F E_k s_j - b_{jk}\|_2^2 + R(\{s_j\}). \qquad (2)$$

We denote the iterative reconstruction method solving Eq. 2 as $\{\hat{s}_j\} = A(\{U_j\}; \{b_{jk}\})$. With this notation, the sampling pattern optimization problem consists of finding, for a given under-sampling ratio $\gamma$ and a given set of fully sampled training data $\{b_{jk}^i, s_j^i\}_{i=1\ldots N}$, the sampling pattern $\{\hat{U}_j\}$ that solves:

$$\{\hat{U}_j\} = \underset{\{U_j\}}{\operatorname{argmin}} \, G(\{U_j\}) = \underset{\{U_j\}}{\operatorname{argmin}} \frac{1}{N} \sum_{i=1}^{N} L(\{\hat{s}_j^i\}, \{s_j^i\}),$$

$$\text{subject to } \{\hat{s}_j^i\} = A(\{U_j\}; \{U_j b_{jk}^i\}) \text{ and } \bar{U}_j = \gamma \text{ for all } i \text{ and } j, \qquad (3)$$

where $N$ is the total number of samples in the training dataset, $\{s_j^i\}$ is the $i$-th fully sampled multi-echo image, $\{\hat{s}_j^i\}$ is the $i$-th reconstructed under-sampled multi-echo obtained using solver

$A(\{U_j\}; \{U_j b_{jk}^i\})$ and $L$ is the metric to quantify difference between $\{\hat{s}_j^i\}$ and $\{s_j^i\}$, such as the $L_1$ loss. In the following section, we will propose a unified framework called LARO (Learned Acquisition and Reconstruction Optimization) to tackle both Eq. 2 and 3 using deep learning techniques.

**Sampling pattern optimization (SPO)**

For k-space sampling pattern optimization Eq. 3, we extend the previously proposed LOUPE-ST method (15) to the multi-echo setting. We consider 2D variable density Cartesian sampling patterns in the $k_y - k_z$ plane with a fixed under-sampling ratio as shown in Figure 1b, in which learnable weights $\{w_j\}$ are used to generate a multi-echo probabilistic pattern $\{P_j\}$ through sigmoid transformation and sampling ratio renormalization:

$$P_j = \text{Renorm}\left(\frac{1}{1 + e^{-a \cdot w_j}}\right), \qquad (4)$$

where $a$ is the slope parameter of the sigmoid function and $\text{Renorm}(\cdot)$ is a linear scaling operation to make sure the mean value of probabilistic pattern is equal to the desired under-sampling ratio (14). Assuming an independent Bernoulli distribution $Ber(P)$ at each k-space location, a binary under-sampling pattern $U_j$ is generated via stochastic sampling from $P_j$:

$$U_j = \mathbf{1}_{z < P_j}, \qquad (5)$$

where $\mathbf{1}_x$ is the indicator function on the truth value of $x$ and $z$ is uniformly distributed between [0, 1]. Then $\{U_j\}$ are used to retrospectively acquire $\{b_{jk}\}$ from fully sampled multi-echo k-space data. The stochastic sampling layer in Eq. 5 has zero gradient almost everywhere when backpropagating through this layer, which makes updating $\{w_j\}$ infeasible (27). To solve this issue, LOUPE-ST implements a straight-through estimator (28) for backpropagation through the stochastic sampling layer by using the probability distribution P instead:

$$\frac{d\mathbf{1}_{z<P_j}}{dw_j} \rightarrow \frac{dP_j}{dw_j}, \tag{6}$$

which solves the zero gradient issue and performs better than other gradient approximations, such as the one implemented in LOUPE (15).

**Temporal feature fusion (TFF) for reconstruction**

For image reconstruction Eq. 2, we propose an unrolled architecture with a temporal feature fusion (TFF) module based on the plug-and-play ADMM (29) strategy. In plug-and-play ADMM, auxiliary variables $v_j = s_j$ for each echo $j$ were introduced and an off-the-shelf image denoiser $\{v_j^{(t+1)}\} = \mathcal{D}\left(\{\tilde{v}_j^{(t)}\}\right)$, where $\tilde{v}_j^{(t)} = s_j^{(t)} + \frac{1}{\rho}u_j^{(t)}$ with $u_j^{(t)}$ the dual variable of the $t$-th outer loop and $\rho$ the penalty parameter in ADMM, was applied. We propose to unroll the iterative scheme of plug-and-play ADMM as a data graph which we call "deep ADMM" network as shown in Figure 1a, where a CNN denoiser $\mathcal{D}\left(\{\tilde{v}_j^{(t)}\}; w_D\right)$ with weights $w_D$ is designed to replace $\mathcal{D}\left(\{\tilde{v}_j^{(t)}\}\right)$ as:

$$v_j^{(t+1)} = \mathcal{D}\left(\tilde{v}_j^{(t)}; w_D\right). \tag{7}$$

To incorporate the dynamic nature of multi-echo images into $\mathcal{D}\left(\{\tilde{v}_j^{(t)}\}; w_D\right)$, we propose a temporal feature fusion (TFF) module as shown in Figure 1c. In TFF, a recurrent module is repeated $N_T$ times in which at the $j$-th repetition (corresponding to the $j$-th echo), $s_j$ (real and imaginary parts concatenated along the channel dimension) and $s_{j-1}$'s hidden state feature $h_{j-1}$ are fed into the module to generate $s_j$'s hidden state feature $h_j$:

$$h_j = ReLU\left(N_s(s_j) + N_h(h_{j-1})\right), \tag{8}$$

where $N_s(\cdot)$ and $N_h(\cdot)$ are convolutional layers for $s_j$ and $h_{j-1}$, and $ReLU$ is the Rectified Linear Unit activation function. The learnable weights in $N_s(\cdot)$ or $N_h(\cdot)$ are shared across recurrent repetitions. At the $j$-th recurrent forward pass shown in Eq. 8, feature maps $h_j$ are generated by aggregating $s_j$ and $h_{j-1}$ through convolutions and nonlinear activations, which implicitly capture the echo dynamics and fuses features from the preceding echoes. After a full recurrent pass over echoes, all feature maps $h_j$ are concatenated along the batch dimension and fed into a denoising network to generate $\{v_j^{(t+1)}\}$. The dynamic nature of the signal over echo times is implicitly captured with the recurrent forward process due to the parameter sharing mechanism which attempts to exploit the relationship between a given echo and all earlier echoes.

**K-space under-sampling sequence design**

The learned k-space sampling patterns $U_j$ were implemented in an mGRE pulse sequence for prospective data acquisition. Gradient pulses along the phase and slice encoding directions were added between consecutive echoes to allow for the modification of k-space sampling locations echo-by-echo during one TR. To avoid large changes in the phase and slice encoding gradients between two echoes, the following k-space ordering strategy was deployed: for each echo $j$, the sampled k-space locations $U_j$ were first divided into multiple ordered segments of equal size based on their angle with respect to the positive $k_y$ axis. Within each segment, k-space locations were ordered based on their distance with respect to the k-space center. Using such k-space ordering strategy, sampled locations will follow a similar trajectory for all echoes, avoiding large changes in the phase and slice encoding gradients from echo to echo during one TR. Illustration of the proposed segmented k-space ordering and pulse sequence design is shown in Figure 2. In this example, number of echoes $N_T = 10$, acceleration factor $R = 8$, $N_y = 206$, $N_z = 80$, $N_s$ (number of segments) = 11, $N_{ind}$ (number of k-space location per segment) = 188 so that $N_s \times N_{ind} = N_y \times N_z / R$. Figure 2a exemplified the sampled ky-kz locations (yellow dots) in current k-space segment (yellow hollow triangles) during a certain TR. Gy and Gz gradients

(blue solid triangles) in Figure 2b are added between two unipolar readouts in Gx to adjust next sampled location in ky-kz plane.

## METHODS

### Data acquisition and preprocessing

Data were acquired following an IRB approved protocol. All images used in this work were de-identified to protect the privacy of human participants.

Fully sampled acquired k-space data

Cartesian fully sampled k-space data were acquired in 13 healthy subjects (3 females, age: 30.7 ± 7.3) using a 3D mGRE sequence on a 3T GE scanner with a 32-channel head coil. Imaging parameters included FA = 15°, FOV = 25.6 cm, $TE_1$ = 1.972 ms, TR = 36 ms, #TE = 10, ΔTE = 3.384 ms, acquisition matrix = 256×206×80 (readout × phase encoding × phase encoding), voxel size = 1×1×2 $mm^3$, BW = 64 kHz. Total scan time was 9:30 mins per subject. 32-coil k-space data of each echo were compressed into 8 virtual coils using a geometric singular value decomposition coil compression algorithm (30). After compression, coil sensitivity maps of each echo were estimated with a reconstruction null space eigenvector decomposition algorithm ESPIRiT (31) using a centric 20×20×20 self-calibration k-space region for each compressed coil. From the fully sampled data, coil combined multi-echo images were computed using the obtained coil sensitivity maps to provide the ground truth labels for both network training and performance comparison. The central 200 locations along the readout direction were selected, which typically included all locations containing brain anatomy. 8/1/4 subjects (1600/200/800 slices) were used as training, validation, and test datasets, respectively.

To demonstrate the generalization ability of LARO, Cartesian fully sampled k-space data were also acquired in one of the healthy test subjects with the following sequence parameter modifications: another flip angle (25°), number of echoes (7 echoes), voxel size (0.75 × 0.75 × 1.5 $mm^3$), a second MRI scanner from the same manufacturer (GE, 12-channel head coil) and a third MRI scanner from another manufacturer (Siemens, 64-channel head coil). Same k-space processing was applied to these data to get compressed 8-coil k-space, coil sensitivity maps and ground truth labels.

Fully sampled synthetic k-space data

To demonstrate LARO's improvement on pathologic reconstruction, supplementary synthetic k-space datasets from healthy subjects, multiple sclerosis (MS) and intracerebral hemorrhage (ICH) patients were simulated, considering unavailability of acquired fully sampled k-space data from patients. Multi-echo complex images of 7 healthy subjects, 4 MS patients and 1 ICH patient were acquired using a 3D mGRE sequence on a 3T GE scanner. Imaging parameters included FA = 15°, FOV = 25.6 cm, $TE_1$ = 6.69 ms, TR = 49 ms, #TE = 10, $\Delta TE$ = 4.06 ms, acquisition matrix = 256×206×68 (readout × phase encoding × phase encoding), voxel size = 1×1×2 mm$^3$, BW = 64 kHz. Synthetic single-coil k-space data was generated through Fourier transform of the complex multi-echo images. Retrospective Cartesian under-sampling was applied on the synthetic k-space data along two phase encoding directions. The central 200 locations along the readout direction were selected, which typically included all locations containing brain anatomy. Data from 6/1 healthy subjects (1200/200 slices) was used as training/validation. Data from the MS (800 slices) and ICH (200 slices) patients was used as two test datasets.

Under-sampled k-space data in both retrospective and prospective studies

For a retrospective study, an acceleration factor $R = 8$ (12.5% under-sampling ratio) was applied on the fully sampled acquired k-space dataset and acceleration factor $R = 4$ (25% under-sampling ratio) was applied on the fully sampled synthetic k-space dataset. For a prospective study, Cartesian under-sampled k-space data was prospectively acquired in 10 healthy test subjects (3 females, age: 28.4 ± 4.1) using a modified 3D mGRE sequence with the same 3T GE scanner and imaging parameters. Different sampling patterns with $R = 8$ were applied during prospective scans and compared. For the optimized k-space sampling pattern, each echo was divided into 11 segments with 188 locations in each segment, resulting in $188 \times 11 = 2068$ k-space locations to sample in total. Corresponding scan time was 1:20 mins. For reference, the default imaging protocol using the same imaging parameters except for elliptical R=2 uniform under-sampling reconstruction using the SENSE implementation (10) on the scanner was performed on the same subjects.

**Implementation details**

Network architecture

The proposed network architecture is shown in Figure 1. Real and imaginary parts of multi-echo images were concatenated along the channel dimension, yielding 20 channels to represent multi-echo complex images in the network. Under-sampled k-space data was zero-filled and Fourier-transformed to be used as input for deep ADMM (Figure 1a) with $N_I = 10$ unrolled iterations. In deep ADMM, the denoiser $\mathcal{D}(\cdot; w_D)$ consisted of five convolutional layers equipped with 320 channels with instance normalization (32) + ReLU activation after convolution for each hidden layer. The TFF module (Figure 1c) used 64 channels in both convolutional layers for $s_j$ and $h_j$. The hidden state feature maps $h_j$ were concatenated along the channel dimension and fed into $\mathcal{D}(\cdot; w_D)$ to generate denoised multi-echo images. The SPO module (Figure 1b) was used to learn optimal sampling patterns, where weights $\{w_j\}$ (with matrix size 206×68×10 for synthetic k-space data and 206×80×10 for the acquired k-space data) were initialized as zeros and slope parameter $a$ in sigmoid function was 0.25. After generating binary patterns $\{U_j\}$ from probabilistic patterns $\{P_j\}$, values in central $20 \times 20$ locations of $\{U_j\}$ were set as ones for self-calibration.

Training strategy

The training process consists of two phases. In phase one, weights in the deep ADMM network and SPO module were updated simultaneously by maximizing a channel-wise structural similarity index measure (SSIM) (33): $\frac{1}{N}\sum_i^N \sum_{j=1}^{N_T} SSIM(\hat{s}_j^i, s_j^i)$ with the measure between two windows $x$ and $y$ of common size ($10 \times 10$) and location in $\hat{s}_j^i$ and $s_j^i$ as:

$$SSIM(x, y) = \frac{(2\mu_x \mu_y + c_1)(2\sigma_{xy} + c_2)}{(\mu_x^2 + \mu_y^2 + c_1)(\sigma_x^2 + \sigma_y^2 + c_2)}, \quad (9)$$

where $\mu_x, \mu_y$ and $\sigma_x, \sigma_y$ are the mean and variance of $x$ and $y$, $\sigma_{xy}$ is the covariance between $x$ and $y$, $c_1 = 0.01^2$ and $c_2 = 0.03^2$. In phase two, the pre-trained deep ADMM network from phase one was fine-tuned with fixed binary sampling patterns $\{U_j\}$ either manually designed using a multi-level sampling scheme (34) or generated from the learned probabilistic patterns $\{P_j\}$ in phase one. We implemented in PyTorch using the Adam optimizer (35) (batch size 1, number of epochs 100 and initial learning rate $10^{-3}$) on a RTX 2080Ti GPU. Our code is available at https://github.com/Jinwei1209/LARO.git.

Ablation study

An ablation study regarding the effectiveness of TFF and SFO modules were investigated by removing one or more of these modules and quantifying the corresponding loss in performance. First, a manually designed variable density sampling pattern was generated based on a multi-level sampling scheme (34) and used to train a baseline deep ADMM network without TFF or SPO (denoted by TFF=0/ SPO=0). Then TFF (denoted as TFF=1), single-echo SPO (optimized sampling pattern was fixed across echoes, denoted as SPO=1) and multi-echo SPO

(denoted as SPO=2) were progressively added to the baseline deep ADMM network to check the effectiveness of each module, with LARO representing TFF with multi-echo SPO (i.e., TFF=1, SPO=2). For baseline deep ADMM without TFF, Eq. 8 was replaced with $h_j = ReLU\left(N_s(s_j)\right)$ by removing $N_h(h_{j-1})$ to show the effectiveness of recurrent forward pass of hidden state features $\{h_j\}$ in TFF, where two 64-channel convolutional layers in $N_s(\cdot)$ were used to match the memory usage of TFF during ablation study.

Performance comparison

Iterative method locally low rank (LLR) (20) and a deep learning method MoDL (21) were used as two benchmark reconstruction methods, where MoDL was modified to reconstruct multi-echo images simultaneously with concatenated real and imaginary parts of multi-echoes along channel dimension. Manually designed and optimized sampling patterns were applied to all reconstruction methods and compared. From the resulting gradient echo images, R2* was estimated using ARLO (36) and QSM using morphology enabled dipole inversion with CSF-0 reference (37) from relative difference local field (RDF), which was estimated using nonlinear field estimation (6), phase unwrapping and background field removal (38).

For all retrospectively under-sampled datasets, quantitative comparisons were presented with fully sampled data as reference, where PSNR (Peak Signal-to-Noise Ratio) and SSIM (Structural Similarity Index) (33) metrics per reconstructed coronal slice were used to measure the reconstruction accuracy of the echo-combined magnitude image $\sqrt{\sum_{j=1}^{N_T}|s_j|^2}$, R2* and RDF maps. RMSE (Root-Mean-Square Error), HFEN (High-Frequency Error Norm) (39) and SSIM (33) per 3D volume were used to measure the reconstruction quality of QSM.

For the MS patient dataset, lesions were manually segmented by an experienced neuroradiologist based on the corresponding T2FLAIR maps which were spatially registered to the magnitude of mGRE data. A linear regression was performed of the mean susceptibility of all lesions between fully sampled and under-sampled test data.

For the prospectively under-sampled dataset, reconstructions were performed by LLR, MoDL and TFF reconstructions with different sampling patterns. The SENSE reconstruction from the scanner with acceleration factor 2 was used as a reference for comparison. Detailed structures in QSM and R2* such as white matter tracts were qualitatively compared. The perivascular spaces were segmented manually into a single region of interest ROIp. From this ROIp, a border ROIb was computed by dilated ROIp by 1 pixel and removing the original ROIp. The sharpness was defined as the difference of average susceptibility of ROIp and ROIb. Mean susceptibility values and standard deviations in manually drawn ROIs including Globus pallidus (GP), Substantia Nigra (SN), Red Nucleus (RN), Caudate Nucleus (CN), Putamen (PU), thalamus (TH), optic radiation (OR) and cerebral cortex (CC, starting from the top of the brain, drawn on the tenth slice of QSMs covering some part of frontal and parietal lobes) were computed and compared.

Generalization experiments

When acquiring the fully sampled test data with sequence parameter modifications, only one parameter was modified in each scan, except for a different voxel size, where increased spatial resolution also increased echo spacing ΔTE to 4.728 ms and acquisition matrix to 320×258×112 (readout × phase encoding × phase encoding). Sampling patterns of this voxel size were obtained by interpolating the pre-trained probabilistic sampling distribution Eq. 4 to the new acquisition matrix and generating sampling patterns using Eq. 5 accordingly. For the test data with 7 echoes, the first 7 sampling patterns were used when applying LARO with SPO=2. Fully sampled data were used as the reference for quantitative comparison in R2*, RDF and QSM, except in magnitude due to signal intensity variations of different scans.

## RESULTS

For abbreviations, "TFF=0" or "1" denotes "with" or "without" temporal feature fusion module; "SPO=0", "1" or "2" denotes "without", "with single-echo", or "with multi-echo" sampling pattern optimization. In terms of reconstruction methods, "TFF" denotes the proposed reconstruction with "TFF=1" under different sampling patterns; "LARO" denotes "TFF" reconstruction specifically under "SPO=2" sampling pattern, i.e., the proposed learned acquisition and reconstruction optimization framework.

### Sampling patterns

Figure 3 shows SPO=2 sampling pattern of the first echo (Echo1) and difference maps between two adjacent echoes ($\Delta_{Echo\#}$) in (a): synthetic k-space data (acceleration factor $R = 4$) and (b): acquired k-space data (acceleration factor $R = 8$). Different k-space sampling patterns were generated from the learned probabilistic patterns per echo, introducing additional incoherency along the temporal dimension.

### Acquired k-space data

Ablation study

Reconstructed magnitude, R2*, RDF and QSM in one representative slice are shown in Figure 4. As TFF and SPO modules were gradually added to the baseline deep ADMM architecture, reconstruction errors ($2^{nd}$, $4^{th}$, $6^{th}$ and $8^{th}$ rows) were progressively reduced in all maps, where LARO (TFF=1, SPO=2) performed the best. Depictions of white matter tracts (insets) in R2* and QSM maps were improved as more modules were added. Quantitative metrics of the ablation study is shown in Table S1. Reconstruction accuracies of four maps were progressively improved as more modules were introduced, where LARO (TFF=1, SPO=2) performed the best.

Performance comparison

Reconstructed magnitude, R2*, RDF and QSM with SPO=2 sampling pattern (Figure 3a) in one representative slice are shown in Figure 5. LLR had larger reconstruction errors with heavy block-like artifacts in RDFs and QSMs compared to MoDL and LARO. Pronounced noise in QSMs and R2* (insets) were showed in MoDL, which were not seen in LARO. Reconstructions with SPO=0 and 1 sampling patterns are shown in Figure S1. Quantitative metrics are shown in Table S2. For each method, reconstruction accuracies of magnitude, R2* and QSM maps were progressively improved from sampling pattern SPO=0, 1 to 2. For each sampling pattern, TFF reconstruction consistently outperformed MoDL and LLR.

**Synthetic k-space data**

Ablation study

Reconstructed magnitude, R2*, RDF and QSM at one representative slice of MS test dataset are shown in Figure S2 with quantitative metrics of ablation study in Table S3. Similar to the acquired k-space data, reconstruction accuracies were progressively improved as more modules were added. In Figure S2, putamen in QSMs (insets in QSMs) were better depicted as more modules were added.

Performance comparison on MS dataset

Reconstructed magnitude, R2*, RDF and QSM with SPO=2 sampling pattern (Figure 3b) in one representative slice are shown in Figure 6. LLR had much larger errors compared to MoDL and TFF. TFF slightly outperformed MoDL. Reconstructions with SPO=0 and 1 sampling patterns are shown in Figure S3. Quantitative metrics are shown in Table S4. Both TFF and SPO=2 outperformed other baseline reconstruction methods and sampling patterns.

Linear regressions of lesion-wise mean susceptibility values between fully sampled and reconstructed QSMs are shown in Figure S4. For SPO=0, 1 and 2, linear coefficients for TFF were 1.08, 0.96, and 0.97 with the highest $R^2$: 0.95, 0.98 and 0.99 compared to LLR and MoDL under each sampling pattern. LLR had linear coefficients 1.13, 0.98, 0.95 with the lowest $R^2$: 0.84, 0.81 and 0.92. MoDL had linear coefficients 1.20, 1.07 and 1.10 with $R^2$ in between: 0.89, 0.94 and 0.95. Both TFF and SPO=2 outperformed other baselines.

Performance comparison on ICH dataset

The pre-trained models were tested on the ICH patient data with acceleration factor $R = 4$ and compared. Reconstructed magnitude, R2*, RDF and QSM in one representative slice containing hemorrhage are shown in Figure S5. LLR had the highest errors among the three methods. MoDL showed some errors (red solid arrows) in QSMs which were not seen in TFF. Quantitative metrics show that both TFF and SPO=2 outperformed their baselines.

**Prospective study**

Prospectively under-sampled scans with acceleration factor $R = 8$ were acquired using the modified sequence (Figure 2) with sampling patterns SPO=0, 1 and 2. TFF reconstructions with different sampling patterns are shown in Figure 7, where SENSE reconstructions with $R = 2$ were used as reference. Depictions of white matter tracts in R2* maps (insets in R2* maps) were progressively improved from SPO=0, 1 to 2. Sharpness scores of perivascular spaces inside putamen (insets in QSMs) were 0.0270, 0.0111, 0.0247 and 0.0411 for SENSE, SPO=0, 1 and 2. LARO achieved comparable image quality with R=2 SENSE reference. LLR, MoDL and LARO reconstructions with SPO=2 sampling pattern (Figure 3a) are shown in Figure 8. LLR had the largest errors with heavy block-like artifacts. LARO outperformed MoDL in the depiction of white matter tracts in R2* maps (insets) and vein structures in QSMs (insets). ROI analysis is shown in Table S5. With R=2 SENSE as reference, under-estimations in SN, RN, CN and CC reconstructed by MoDL and TFF were observed when SPO=0 and 1 but were reduced or recovered when SPO=2. LLR had more deviations than MoDL and TFF.

**Generalization Study**

Reconstructions of different test datasets retrospectively under-sampled by SPO=2 were shown in Figure 9. Error maps and quantitative metrics were computed in R2*, RDF and QSM according to their fully sampled references except in magnitude due to signal intensity variations of different datasets. No visible artifacts were seen when applying the pre-trained reconstruction network to the datasets with another flip angle (25°, 2$^{nd}$ column), number of echoes (7 echoes, 1$^{st}$ column) and a second MRI scanner from the same manufacturer (GE, 3$^{rd}$ column). Moderate noise appeared (red arrows in the last column) when tested with another voxel size ($0.75 \times 0.75 \times 1.5\ mm^3$, last column), while heavily pronounced residual aliasing artifacts existed when tested with a third MRI scanner from another manufacturer (Siemens, 4$^{th}$ column). Reconstructions retrospectively under-sampled by SPO=0 and 1 were shown in Figures S6 and S7. For each test dataset, reconstruction performance was consistently improved from sampling pattern SPO=0, 1 to 2.

**DISCUSSION**

In this work, we demonstrated the feasibility of learning a sampling pattern and reconstruction process specifically designed to accelerate the acquisition of multi-echo gradient echo data for the purpose of computing a susceptibility map (QSM). R=8 acceleration was achieved while maintaining QSM quality in both healthy subjects as well as in an MS patient. Both retrospective and prospective acceleration was demonstrated. Finally, reconstruction performance was observed to be superior when compared to previously proposed acceleration techniques.

The original LOUPE (14)/LOUPE-ST (15) learned an optimized variable density sampling pattern from fully sampled single-echo k-space data, which corresponded to SPO = 1 in our work. Since sampling patterns of different variable densities may result in different reconstruction performance, LOUPE/LOUPE-ST updated and optimized such variable density in a learning-based approach. In LOUPE/LOUPE-ST, a variable density sampling pattern was generated from a learnable probabilistic sampling pattern distribution *P* in Eq. 4 that was updated during training to improve the reconstruction performance of under-sampled k-space data. As a result, the optimized variable density from LOUPE/LOUPE-ST outperformed other variable densities, such as manually designed ones.

In this work, multi-echo sampling pattern optimization SPO = 2 (Eq. 3) was learned, achieving both optimized k-space variable density as in SPO = 1 and additional incoherency along echoes, which may result in better aliasing patterns for gradient echo images of different echoes that can be combined and compensated during reconstruction. SPO=2 sampling pattern distinguishes the proposed framework from another deep learning based mGRE acceleration method (24), where manually designed 2D variable density sampling pattern (SPO=0) was applied, which may not be optimal for mGRE acquisition. We extend our conference paper (26) by implementing SPO = 2 sampling pattern into the existing mGRE sequence. The proposed multi-echo adaptive fan-beam ordered strategy (Figure 2a) prevented large changes in the phase and slice encodings between echoes within one TR, improving image quality (40,41). The prospective results in Figures 6 and 7 show the feasibility of achieving R = 8 factor acceleration using the modified mGRE sequence with QSM image quality comparable to R=2 SENSE.

Our reconstruction architecture (Figure 1) was based on unrolling a plug-and-play ADMM iterative scheme (29) and replacing the regularization step with a deep neural network denoiser. This idea is inspired by MoDL (21) where quasi-Newton iterative scheme was unrolled as a network architecture and a five-convolution-layer neural network denoiser was applied. In (24), a MoDL-like architecture (Figure 1 in (24)) was proposed but only one repetition of unrolling was applied. As reported in MoDL (21), more iterations/repetitions of the unrolled architecture helped improve reconstruction performance. We used $N_I = 10$ unrolled iterations same as MoDL to ensure good performance.

Recently, using convolutional neural networks to solve inverse problems related to multi-echo MRI signals has been explored in (42-49), where the established U-Net architecture (50) was always applied. LARO is novel here because it introduces a TFF module (Figure 1c) to implicitly capture the multi-echo correlation and effectively compensate temporally incoherent aliasing patterns of the GRE echo signals when SPO = 2. The benefit of the TFF module was apparent in our ablation study (Figures 4 and S2, Tables S1 and S3) and comparison to MoDL (Figures 7, S1, S3 and S5). This distinguishes the proposed framework from (24) as well, since in (24) multi-echo images were only concatenated into channel dimension for convolution.

Pathologies such as hemorrhagic lesions which are not seen in the healthy training data can still be effectively reconstructed by LARO and MoDL with low reconstruction error (2nd row in Figure S5). We speculate that the use of the data consistency module in the proposed method allows for accurate image reconstruction of pathologies not seen during training. Generalization experiments of LARO (Figures 8, S6 and S7) imply that changing the flip angle, number of echoes or MRI scanner from the same manufacturer had small deviation from the training dataset so that the pre-trained reconstruction network generalized well on these test cases, while smaller voxel size containing new features not encountered during training increased generalization errors (red arrows in the last column of Figure 9). Similarly, test data from the MRI system of another manufacturer increased generalization errors (4th column in Figure 9). Despite this limitation, pre-trained sampling patterns from SPO=0, 1 to 2 consistently improved the reconstruction performance on all test datasets, which implies that for brain mGRE acquisition, the optimized k-space variable density distribution (Eq. 4) may be independent of the scanning parameters/manufacturers and can be generalized effectively. LARO is also independent of the number of receiver coil channels used for scan, as both TFF and denoiser networks are applied to the coil-combined image, which also improves the generalization ability of LARO.

For raw k-space data, fully sampled training dataset was only available on healthy volunteers because of long scan time (9:30 mins), which was not feasible on patients. To incorporate patients' dataset for training, an unrolled reconstruction network may be trained without fully sampled k-space data using self-supervised learning (51), where during training, one portion of the under-sampled k-space data is included in the data consistency module and the remaining k-space data is used in a forward model loss, which promises to achieve test results comparable to supervised training on fully sampled data. The reconstruction network of LARO may be enhanced by incorporating under-sampled patient data with such self-supervised learning strategy.

LARO is applied here to mGRE for accelerating QSM that is useful for studying tissue magnetism (52), particularly paramagnetic iron (53), and is promising for assessing various diseases (54), such as multiple sclerosis (55). The proposed combination of sampling and reconstruction optimization can be extended to other mGRE tasks with different organs, such as liver and cardiac QSM (56-58), or other quantitative imaging tasks, such as T1 (1) and T2 (2) mapping, where signal models based on Bloch equations are used to describe signal intensity changes over time. The proposed sampling strategy and temporal feature fusion may be useful to obtain better multi-contrast images. Furthermore, with the emergence of quantitative multi-parametric MRI (59), sampling and reconstructing multi-contrast images together in one sequence can be an effective strategy, since multi-contrast images that are intrinsically registered in one scan have redundancy in both spatial and temporal dimensions, which can be utilized to regularize the image series during reconstruction. Our future work will extend LARO to other mGRE and multi-contrast MRI tasks.

**CONCLUSION**

We propose LARO, a unified method to optimize the mGRE signal acquisition and image reconstruction to accelerate QSM. The proposed reconstruction network inserts a recurrent network module into a deep ADMM network to capture the signal evolution and compensate the aliasing artifacts along echo time. The proposed sampling pattern optimization module allows acquiring k-space data along echoes with an optimized multi-echo sampling pattern. Experimental results showed superior performance LARO with good generalization ability. Prospective scan using the optimized multi-echo sampling pattern shows the feasibility of LARO.

# FIGURES

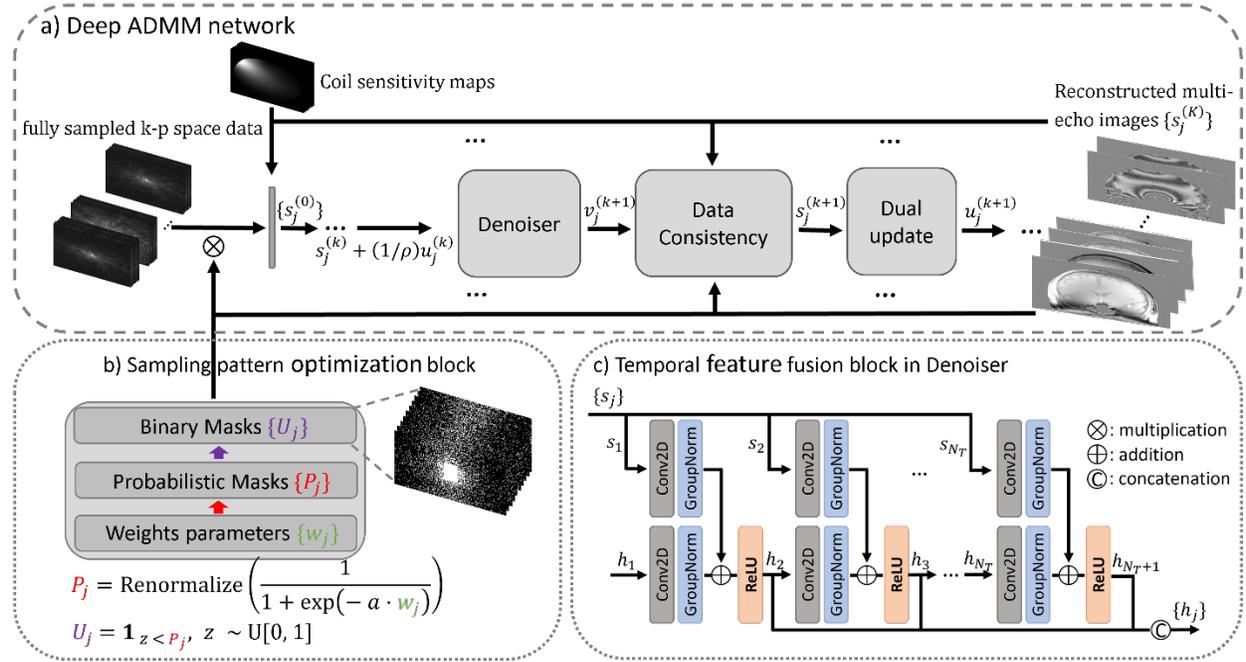

Figure 1. Network architecture of LARO. (a): deep ADMM was used as the backbone for under-sampled k-space reconstruction. (b): a sampling pattern optimization (SPO) module was used to learn the optimal k-space under-sampling pattern. (c): a temporal feature fusion (TFF) module was inserted into deep ADMM to capture the signal evolution along echoes.

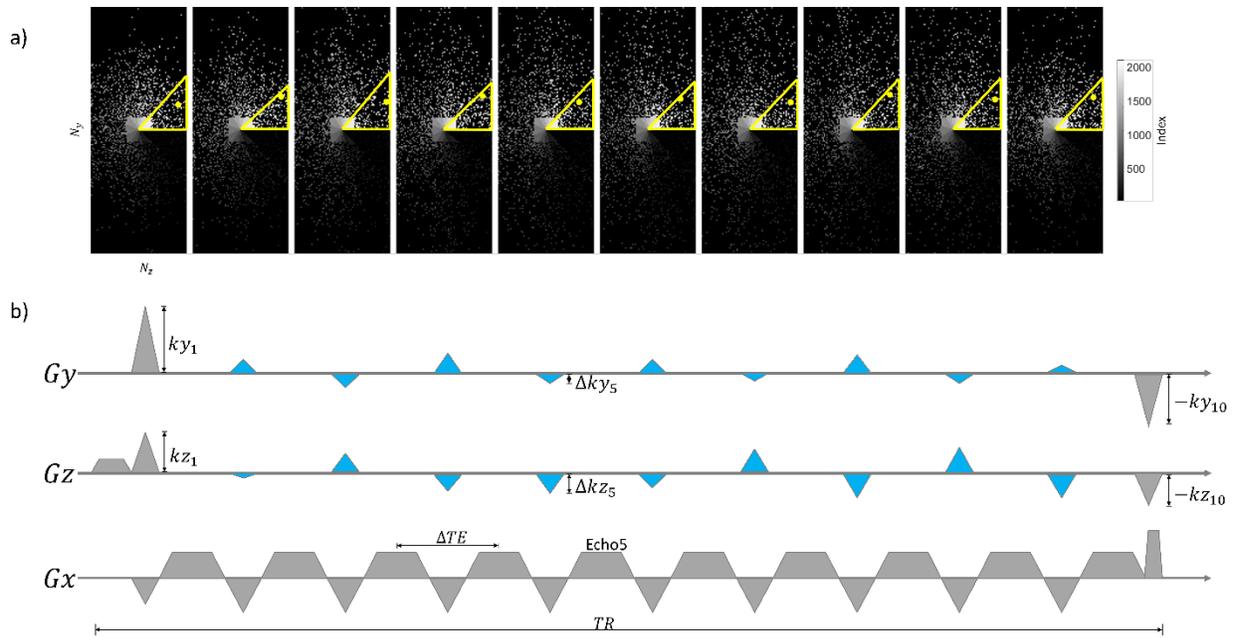

Figure 2. Illustration of (a): the proposed segmented k-space ordering strategy of ten echoes and (b): pulse sequence design. In (a), segmented centric k-space ordering is indexed by greyscale level. In a certain TR, sampled ky-kz locations (yellow dots) in current k-space segment (yellow hollow triangles) are exemplified. In (b), additional Gy and Gz gradients (blue solid triangles) are added between two unipolar readouts in Gx to adjust next sampled location in ky-kz plane.

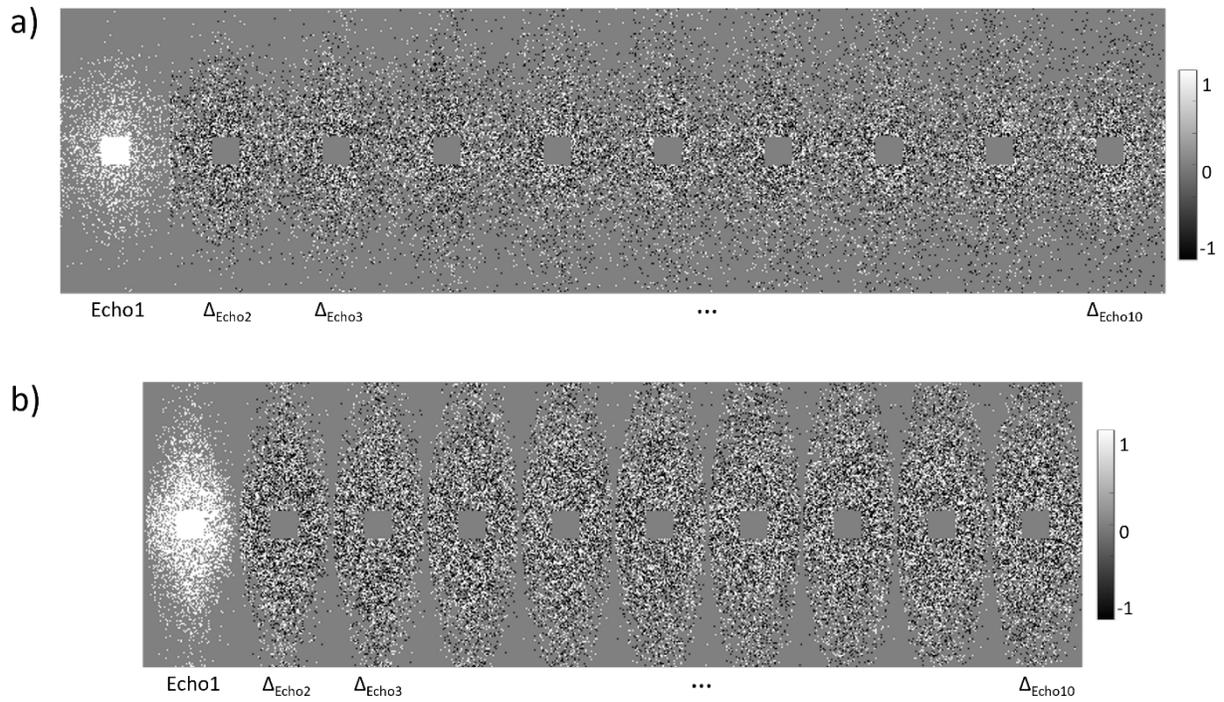

Figure 3. SPO=2 sampling pattern of the first echo (Echo1) and difference maps between two adjacent echoes ($\Delta_{Echo\#}$) in (a): acquired k-space data (acceleration factor $R = 8$) and (b): synthetic k-space data (acceleration factor $R = 4$). Different k-space sampling pattern was generated from the learned probabilistic pattern per echo, introducing additional incoherency along temporal dimension.

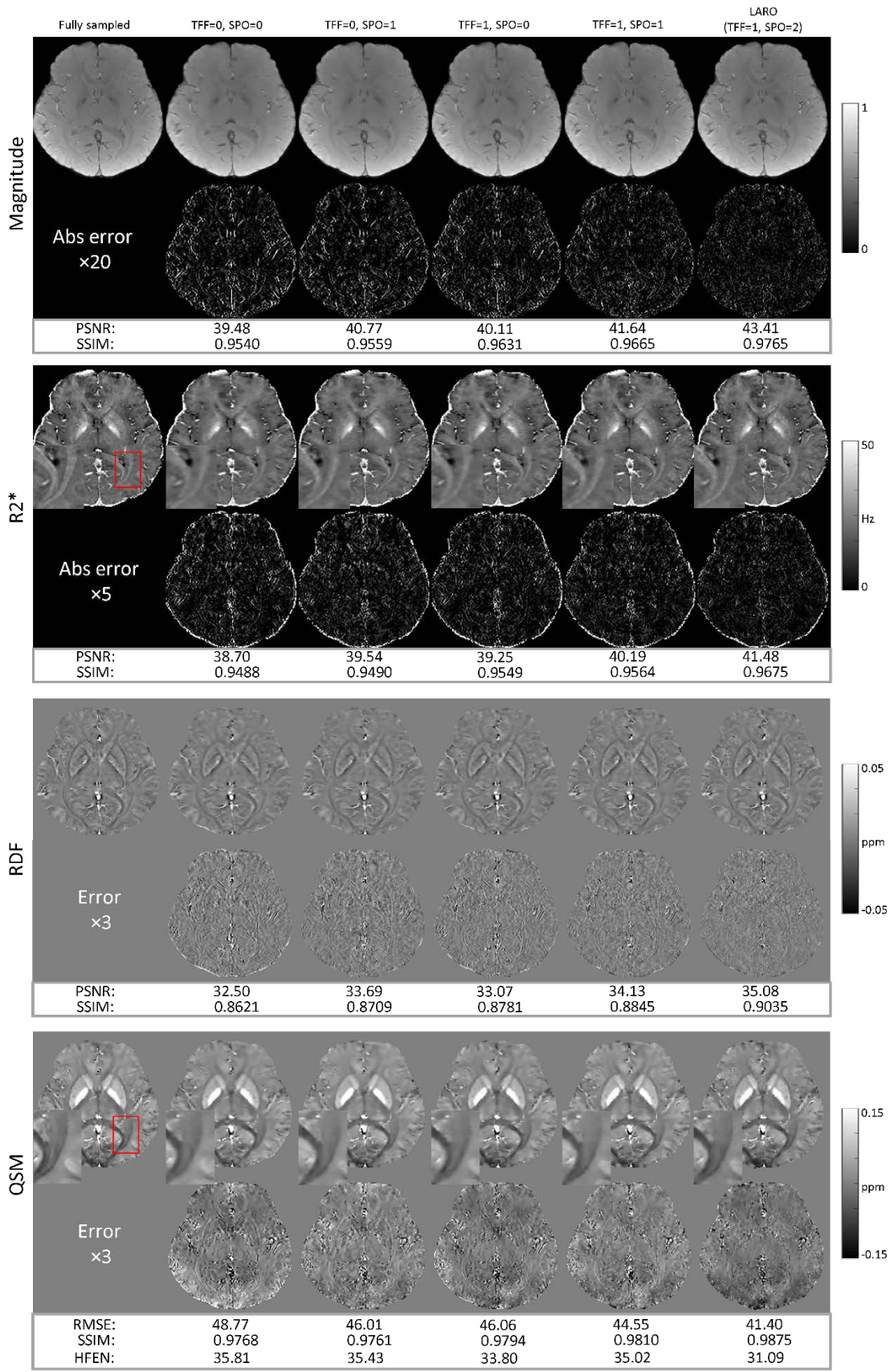

Figure 4. Ablation study on acquired k-space dataset with acceleration factor $R = 8$. Reconstruction errors were progressively reduced in magnitude, R2* and QSM as more modules were added. White matter tracts (insets) were blurry in all reconstructed R2* and QSMs except LARO (TFF=1, SPO=2). Abbreviation: TFF=0 or 1, with or without temporal feature fusion module; SPO=0, 1 or 2, "without", "with single-echo", or "with multi-echo" sampling pattern optimization.

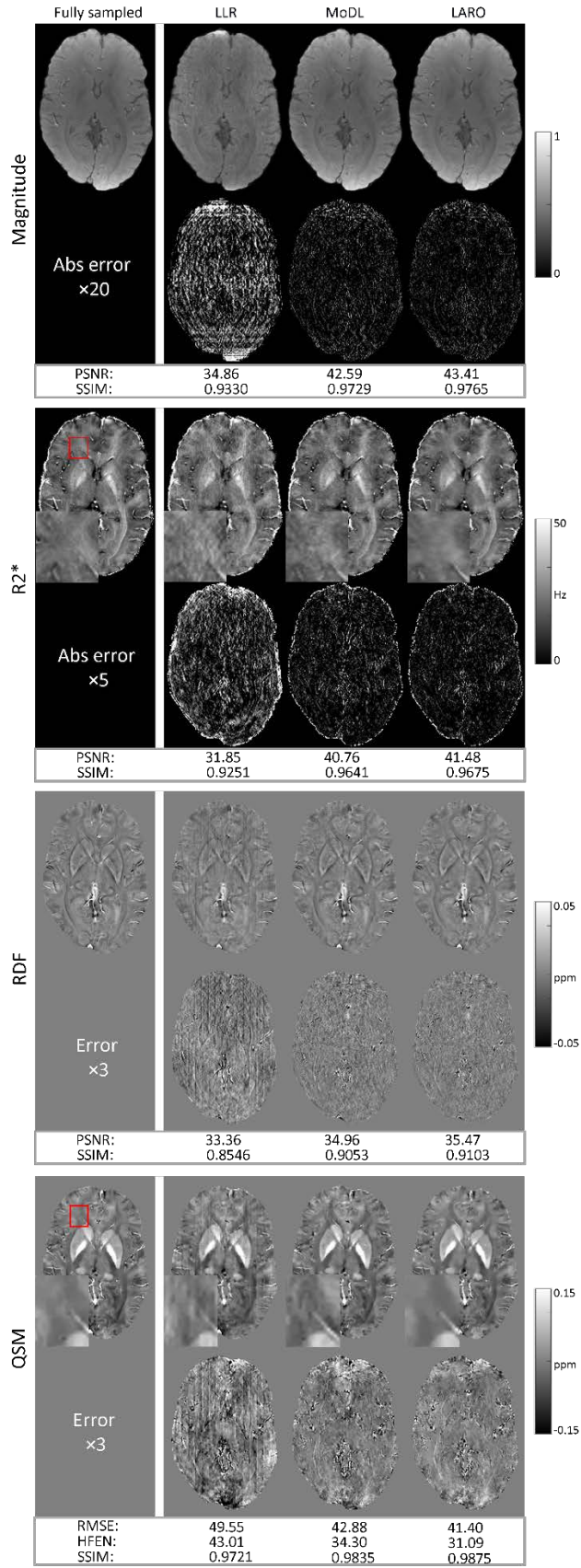

Figure 5. Performance comparison of acquired k-space test dataset under-sampled by the optimized sampling pattern with acceleration factor $R = 8$ (Figure 3a). LLR had heavy block-like artifacts in RDFs and QSMs with larger errors compared to MoDL and LARO. Insets in QSMs and R2* showed pronounced noise in MoDL, which were not seen in LARO.

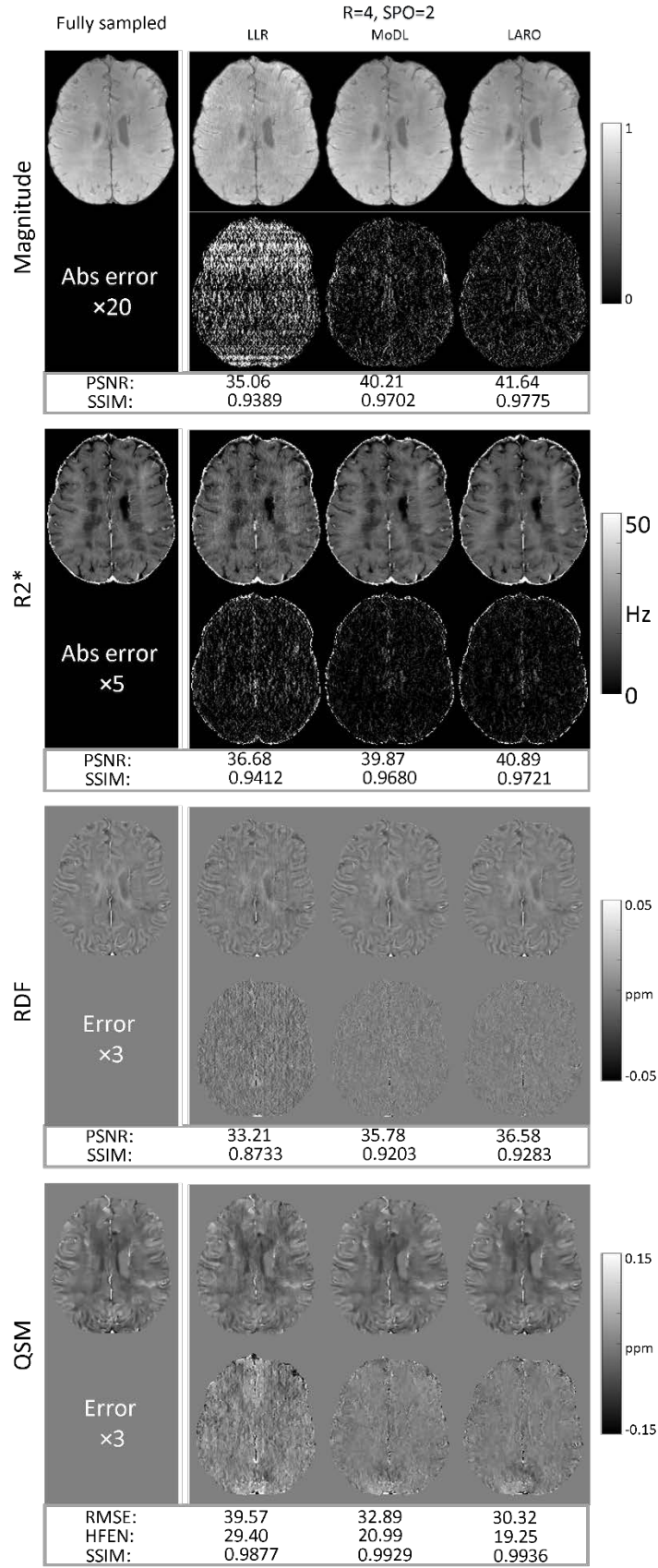

Figure 6. Performance comparison of MS lesion dataset under-sampled by the optimized sampling pattern with acceleration factor $R = 4$ (Figure 3b). MoDL and LARO dramatically outperformed LLR in terms of reconstruction accuracy, while LARO was slightly better than MoDL.

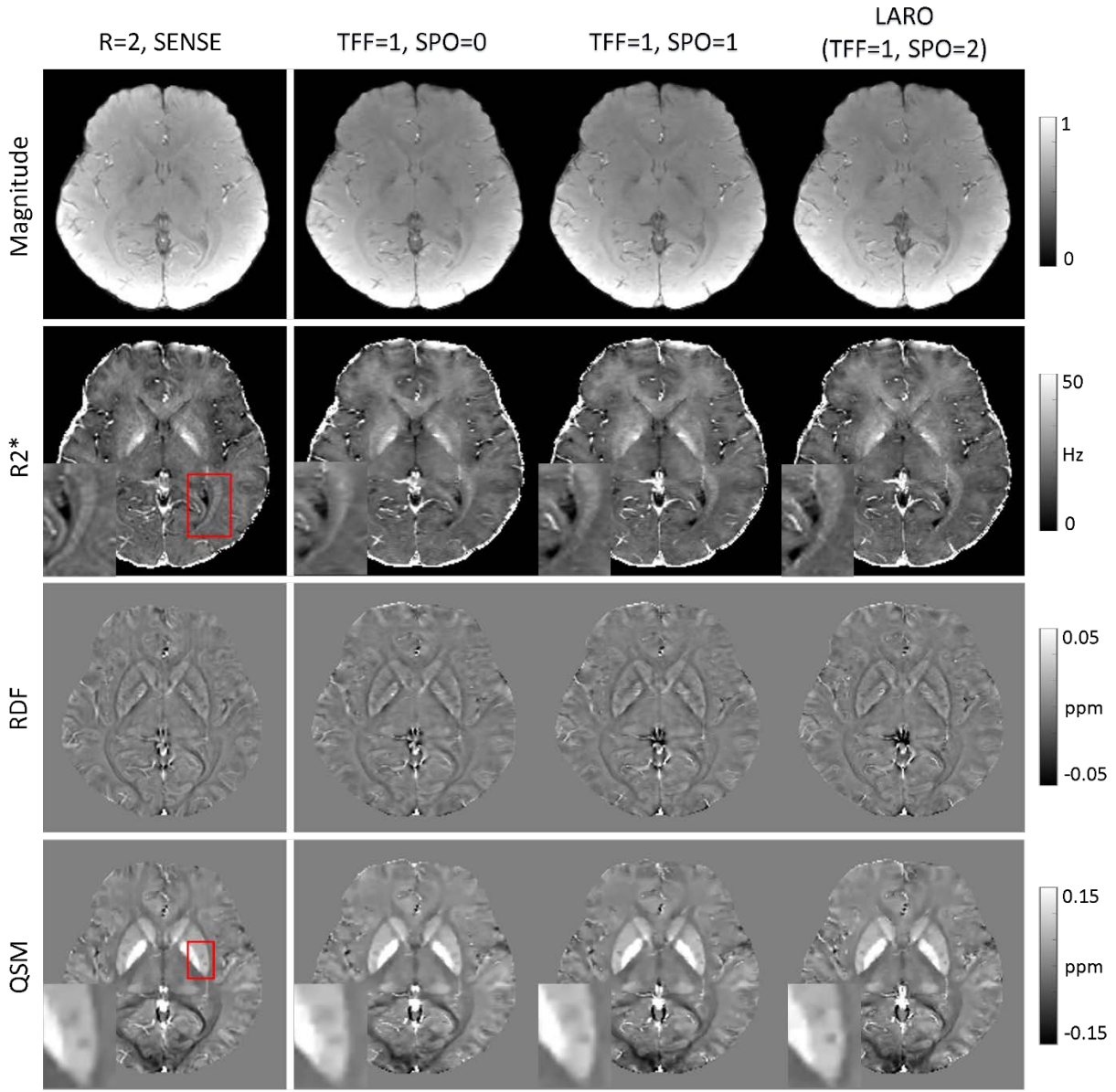

Figure 7. TFF reconstructions on prospectively under-sampled raw k-space data of one healthy subject with acceleration factor $R = 8$. Compared to SENSE reconstruction with $R = 2$ as reference, depictions of white matter tracts in R2* maps (insets in R2* maps) were progressively improved from SPO=0, 1 to LARO (SPO=2). Sharpness scores of perivascular spaces inside putamen (insets in QSMs) were 0.0270, 0.0111, 0.0247 and 0.0411 for SENSE, SPO=0, 1 and 2.

Abbreviation: TFF= 1, with temporal feature fusion module; SPO=0, 1 or 2, without, with single-echo or with multi-echo sampling pattern optimization.

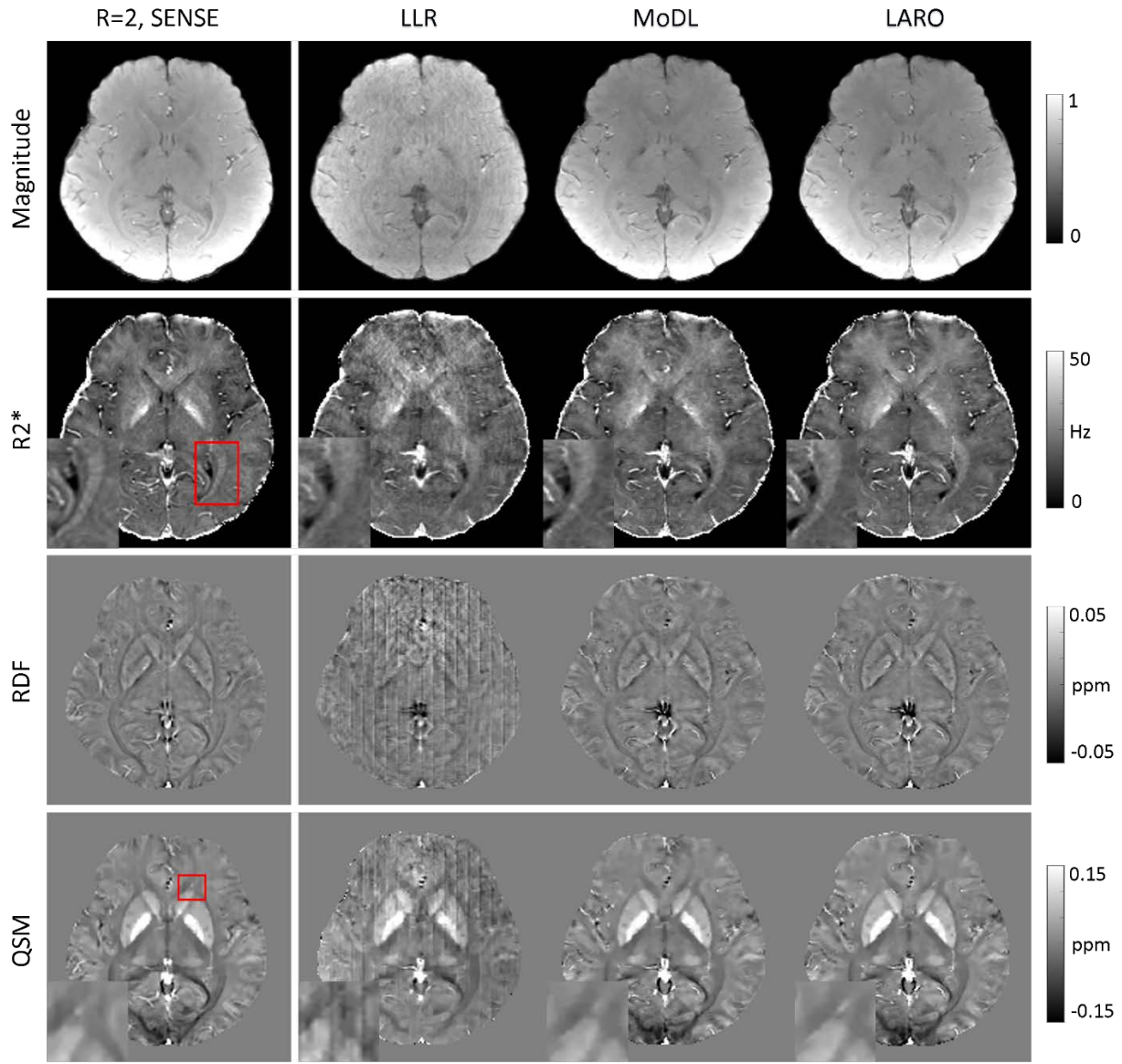

Figure 8. Performance comparison on the prospectively under-sampled raw k-space data of one healthy subject with SPO=2 and acceleration factor $R = 8$ (Figure 3a). SENSE reconstructions with $R = 2$ were used as references. LLR had heavy block-like artifacts in RDFs and QSMs. White matter tracts in R2* maps (insets in R2* maps) and vein structures in QSMs (insets in QSMs) were blurrier in MoDL than LARO.

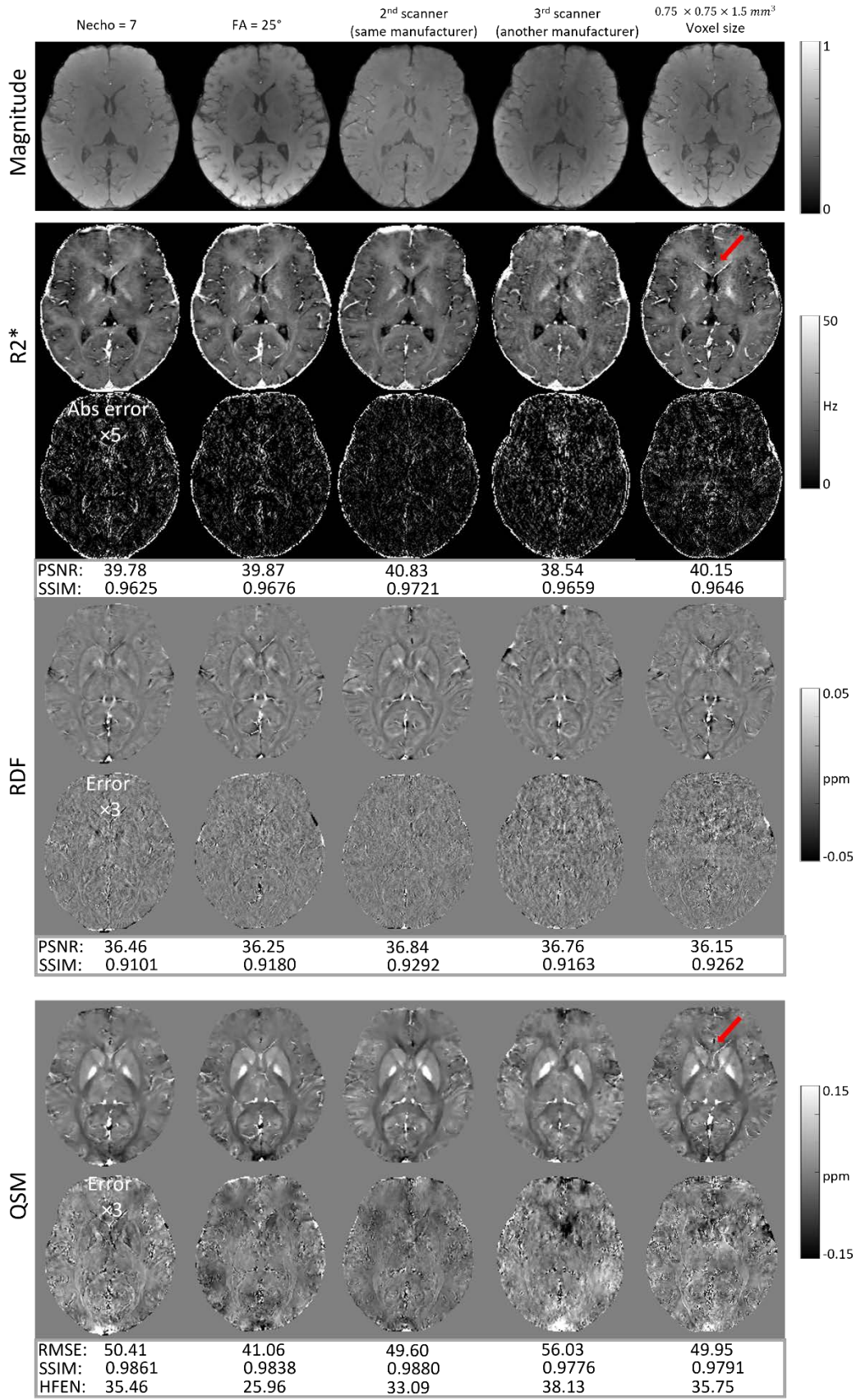

Figure 9. Generalization experiments of LARO with different imaging parameters retrospectively under-sampled by SPO=2 sampling pattern. Fully sampled reference of each test dataset was used to compute error maps and quantitative metrics. Magnitude images were not considered for quantitative comparison due to signal intensity variations among scans. LARO performed well without visible artifacts on test datasets with another flip angle (25°, 2nd column), number of echoes (7 echoes, 1st column) and a second MRI scanner from the same manufacturer (GE, 3rd column), but had moderate noise (red arrows in the last column) on another voxel size ($0.75 \times 0.75 \times 1.5\ mm^3$, last column) and heavily pronounced residual aliasing artifacts on a third MRI scanner from another manufacturer (Siemens, 4th column). Reconstructions on these datasets retrospectively under-sampled by SPO=1 and 0 were shown in Figures S6 and S7. For each test dataset, reconstruction performance was consistently improved from sampling pattern SPO=0, 1 to 2.

**SUPPLEMENTARY INFORMATION**

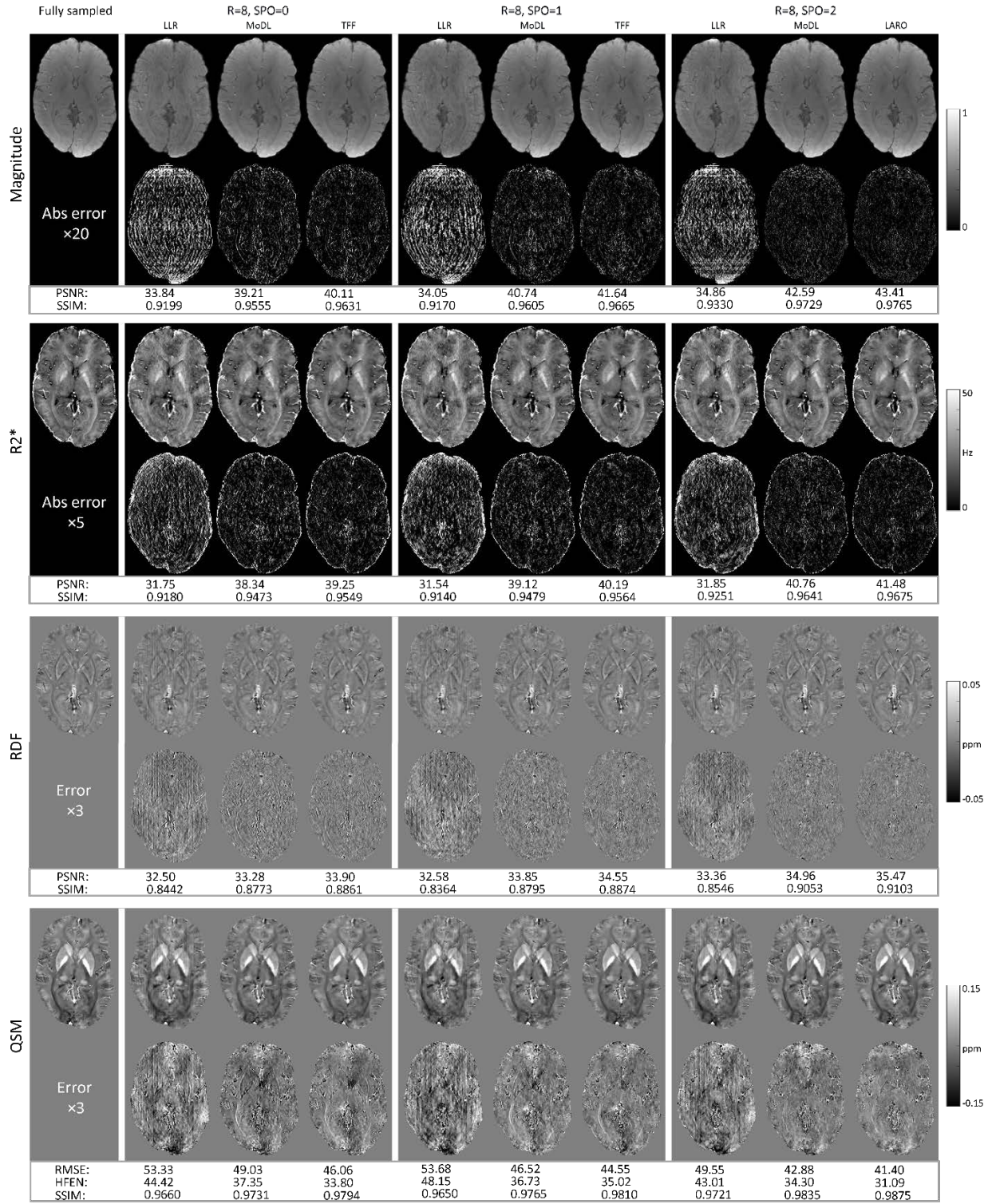

Figure S1. Performance comparison on raw k-space test dataset under-sampled by different patterns with acceleration factor $R = 8$. For each sampling pattern, LLR had much more pronounced noise and larger errors compared to MoDL and TFF. For MoDL and TFF, reconstruction accuracies of four maps were progressively improved from sampling pattern SPO=0, 1 to 2.

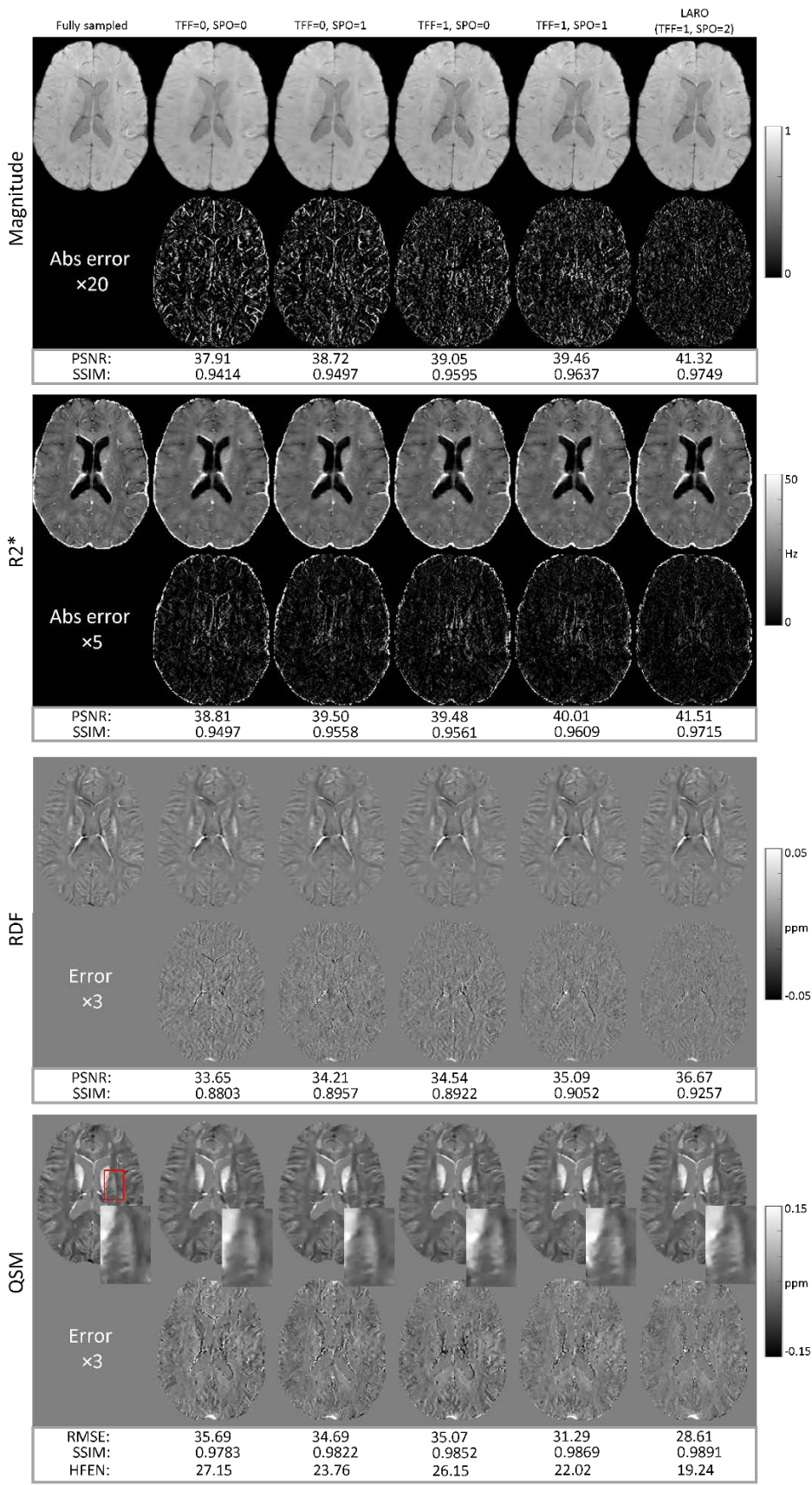

Figure S2. Ablation study on MS lesion dataset with acceleration factor $R = 4$. Reconstruction errors were progressively reduced in all maps as TFF and SPO modules were gradually added to the deep ADMM network. Putamen in QSMs (insets in QSMs) were progressively improved as more modules were added.

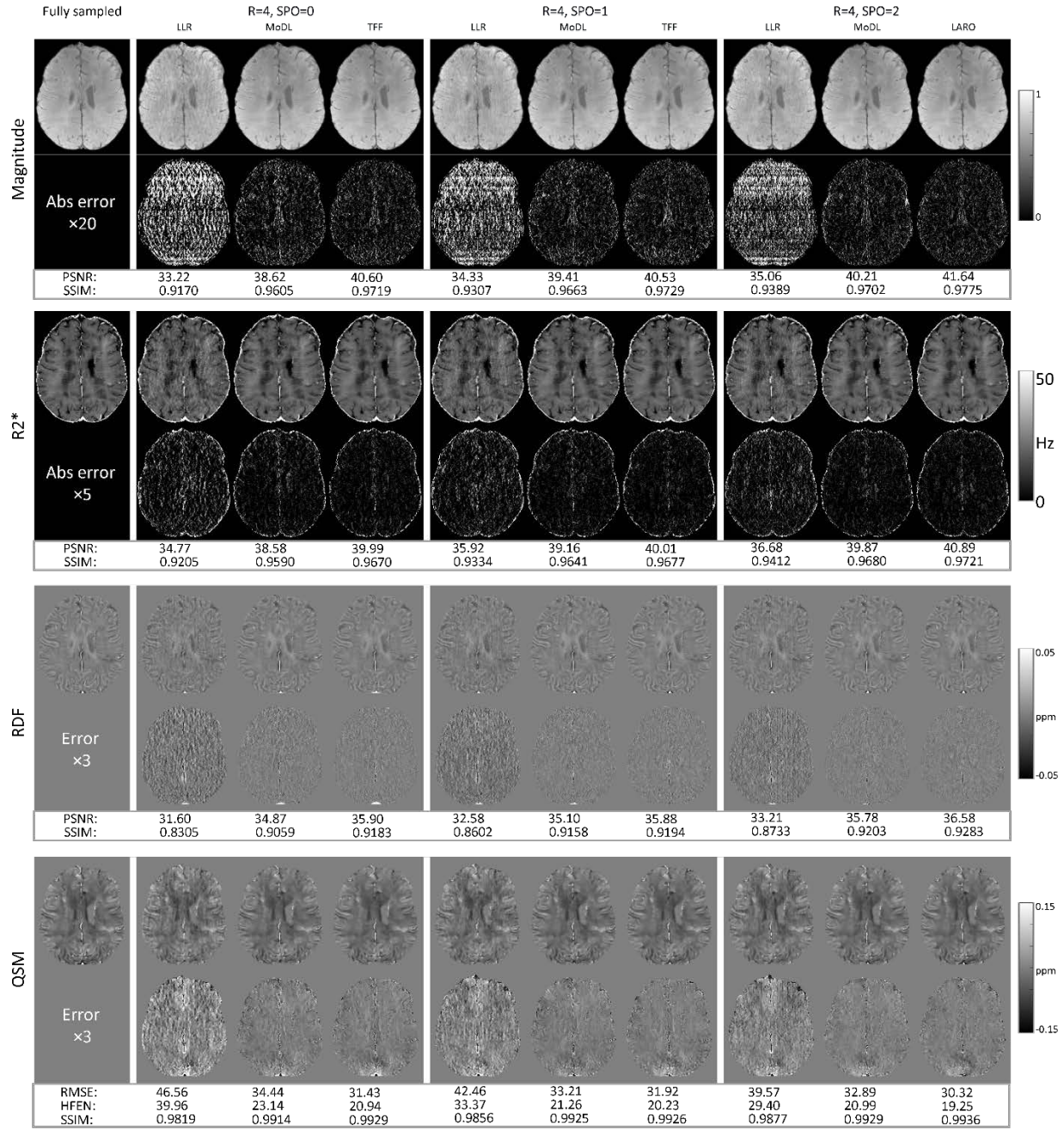

Figure S3. Performance comparison on MS lesion dataset with acceleration factor $R = 4$. TFF consistently outperformed MoDL and LLR for each sampling pattern, where LLR had heavy block-like artifacts in RDFs and QSMs. For MoDL and TFF, reconstruction accuracies of four maps were progressively improved from sampling pattern SPO=0, 1 to 2.

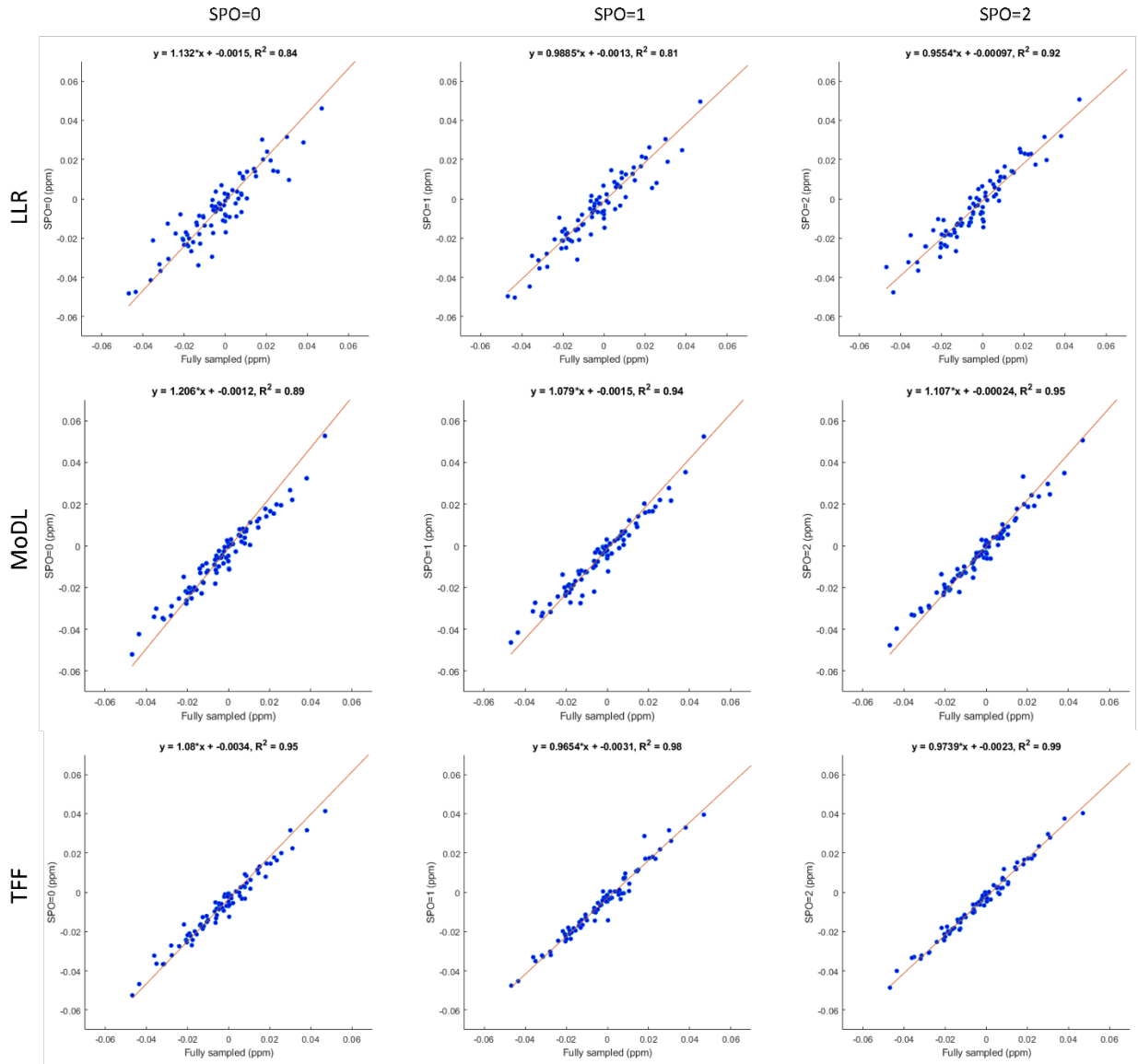

Figure S4. Linear regressions of lesion-wise mean values between fully sampled and different reconstructed QSMs under SPO=0, 1 and 2 sampling patterns. Under each sampling pattern, $R^2$ was progressively improved from LLR, MoDL to TFF. Linear coefficient of TFF was closer to 1 than that of LLR and MoDL under SPO=0 and 2, while under SPO=1 LLR showed linear coefficient closer to 1 than TFF but with lower $R^2$. Under each reconstruction method, SPO=2 pattern always showed better $R^2$ than SPO=0 and 1 patterns. Overall, TFF with SPO=2 pattern performed the best in terms of both $R^2$ and linear coefficient.

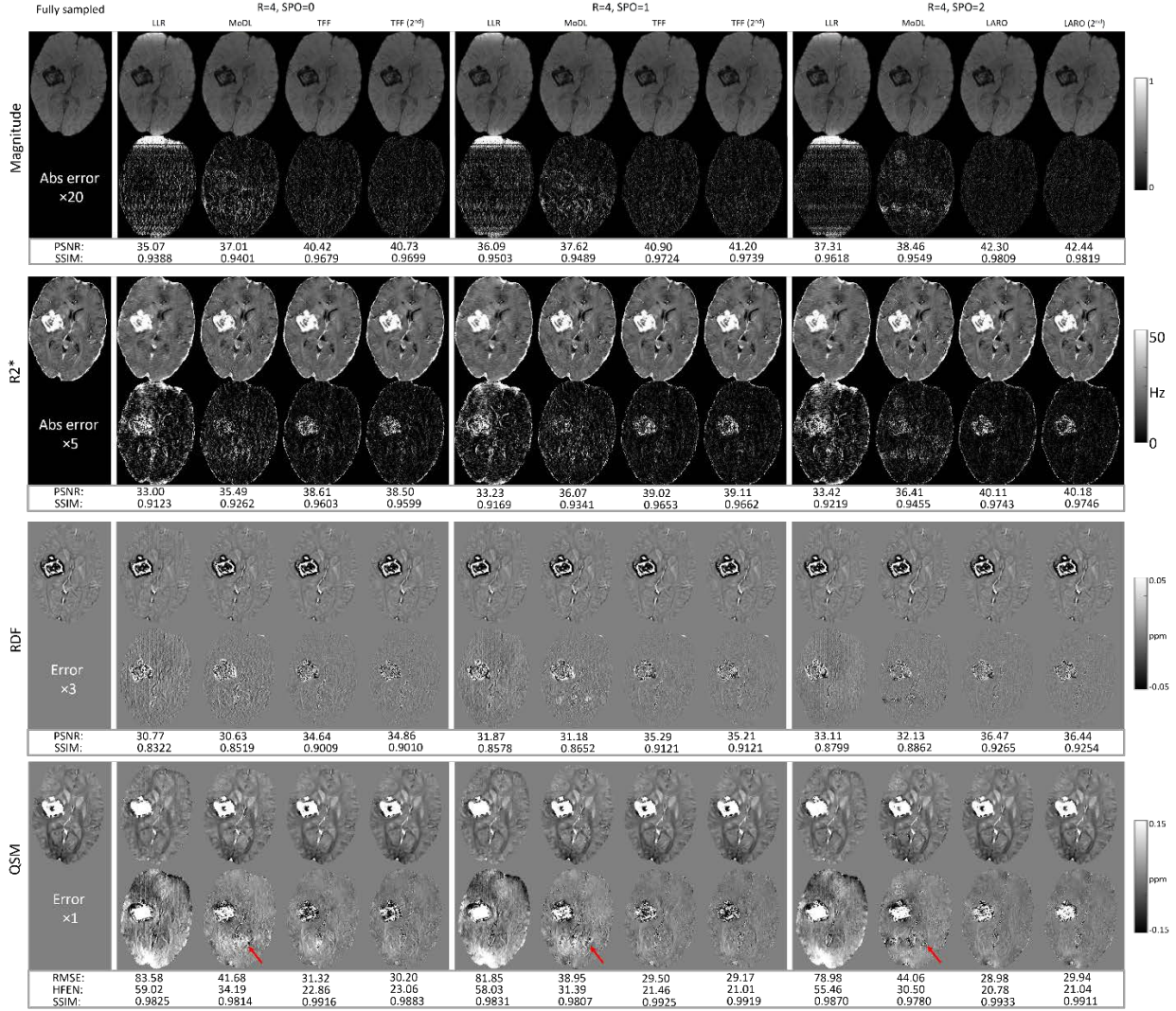

Figure S5. Performance comparison on ICH data with acceleration factor $R = 4$. LLR had the highest errors among the three methods. Reconstruction accuracies of four maps were progressively improved from sampling pattern SPO=0, 1 to 2 TFF. TFF consistently outperformed MoDL and LLR for each sampling pattern. MoDL showed some reconstruction errors not seen in TFF for QSMs (red solid arrows).

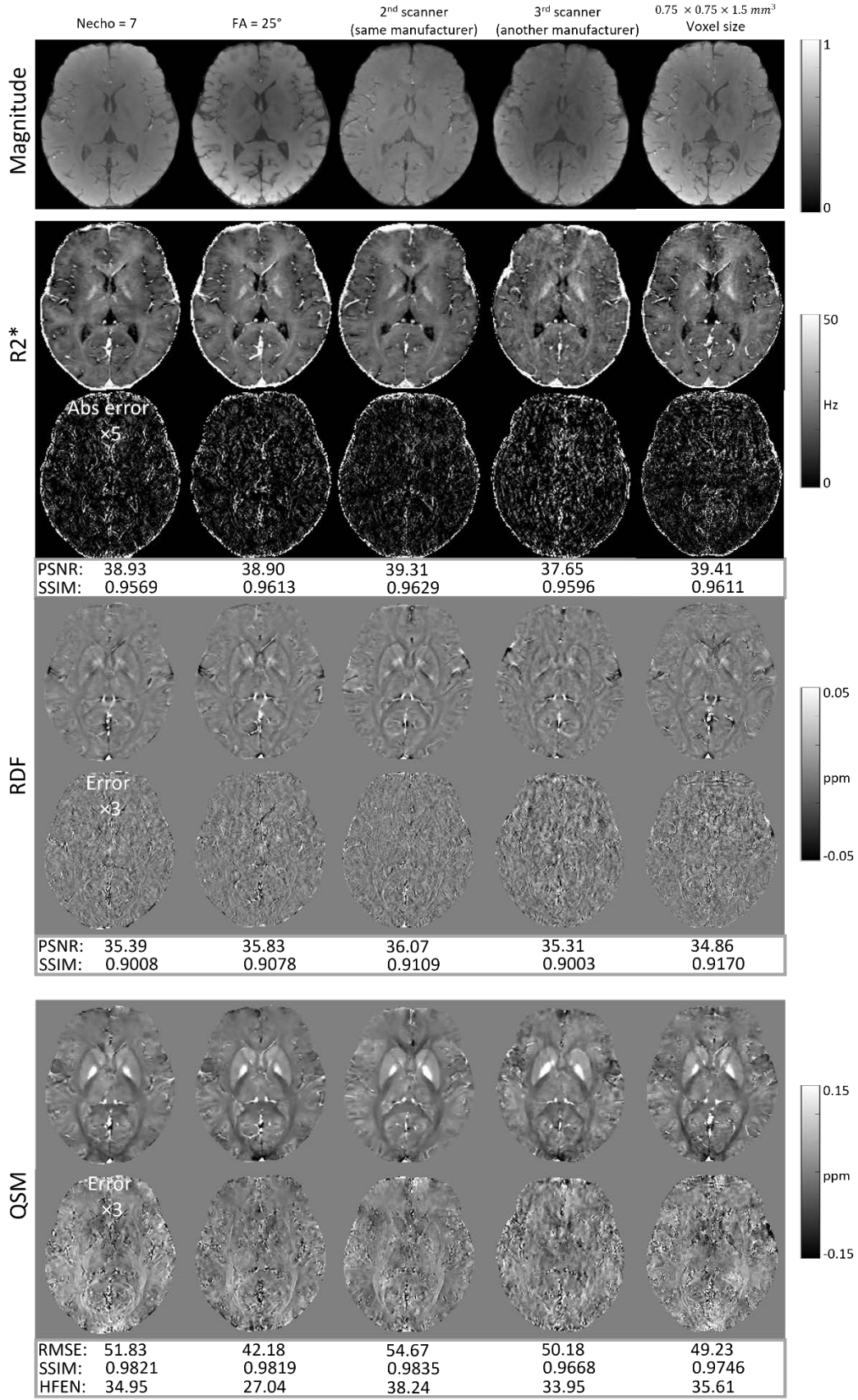

Figure S6. Generalization experiments of LARO with different imaging parameters retrospectively under-sampled by SPO=1 sampling pattern (supplementary to Figure 9). Similar to SPO=2 in Figure 9, LARO worked well on test datasets with another flip angle (25°, 2nd column), number of echoes (7 echoes, 1st column) and a second MRI scanner from the same manufacturer (GE, 3rd column), but had moderate noise on another voxel size (0.75 × 0.75 × 1.5 $mm^3$, last column) and heavily pronounced residual aliasing artifacts on a third MRI scanner from another manufacturer (Siemens, 4th column). For each dataset, reconstruction performance was consistently degraded compared to SPO=2 in Figure 9.

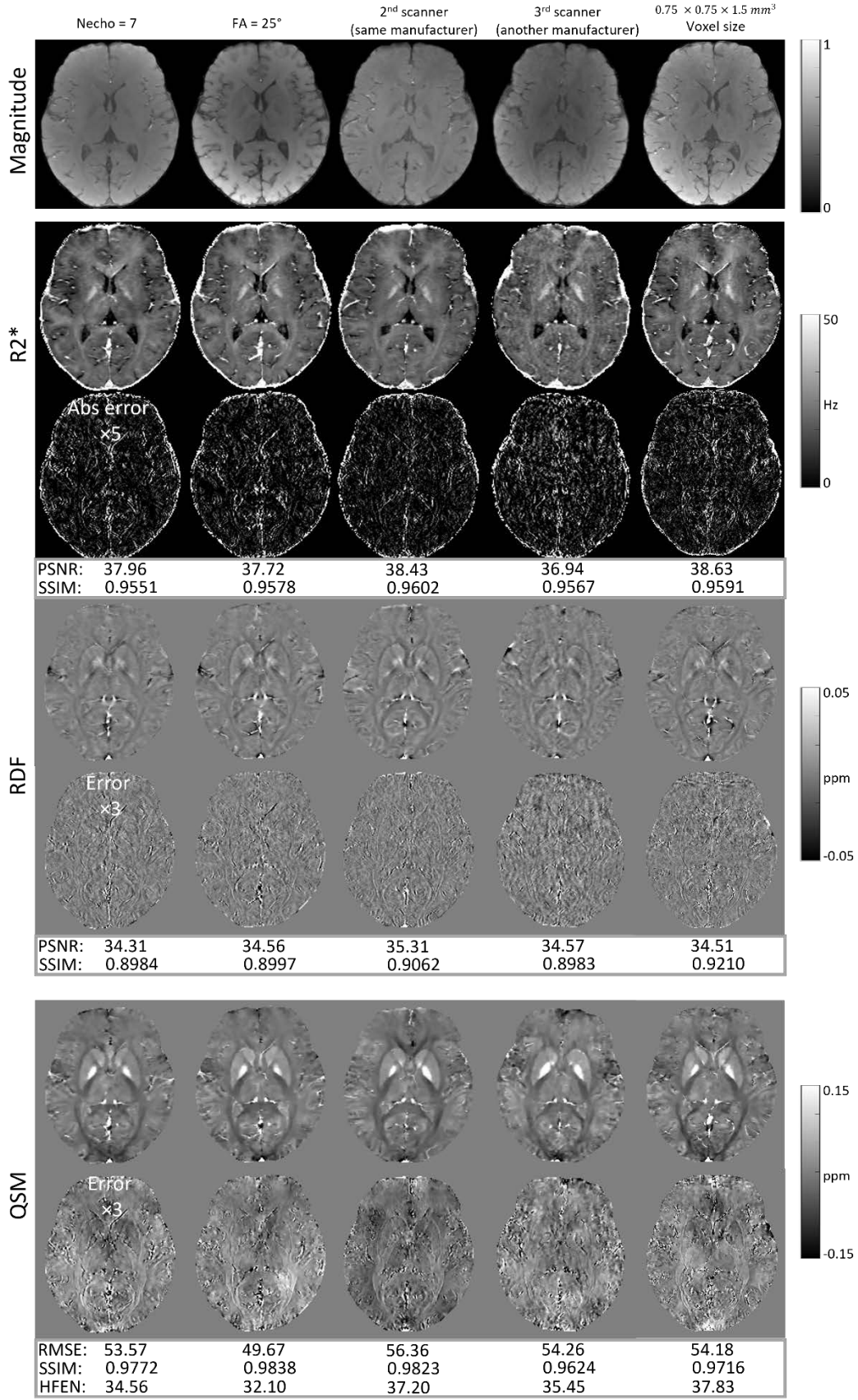

Figure S7. Generalization experiments of LARO with different imaging parameters retrospectively under-sampled by SPO=0 sampling pattern (supplementary to Figure 9). Similar to SPO=2 and 1 in Figures 8 and S6, LARO worked well on test datasets with another flip angle (25°, 2nd column), number of echoes (7 echoes, 1st column) and a second MRI scanner from the same manufacturer (GE, 3rd column), but had moderate noise on another voxel size ($0.75 \times 0.75 \times 1.5\ mm^3$, last column) and heavily pronounced residual aliasing artifacts on a third MRI scanner from another manufacturer (Siemens, 4th column). For each dataset, reconstruction performance was consistently degraded from SPO=2 in Figure 9, SPO=1 in Figure S6 to SPO=0.

Table S1. Ablation study on retrospectively under-sampled raw k-space data (800 2D coronal slices of magnitude, R2* and RDF, and 4 3D volumes of QSM) with acceleration factor $R = 8$. Reconstruction accuracies (mean ± standard) of four maps were progressively improved as more modules were added (*: statistical significance compared to LARO, p < 0.05).

| Magnitude | PSNR (↑) | SSIM (↑) |
|---|---|---|
| TFF=0, SPO=0 | 39.48 ± 2.93* | 0.9524 ± 0.0258* |
| TFF=0, SPO=1 | 40.77 ± 2.97* | 0.9548 ± 0.0250* |
| TFF=1, SPO=0 | 40.13 ± 3.14* | 0.9616 ± 0.021* |
| TFF=1, SPO=1 | 41.65 ± 2.89* | 0.9656 ± 0.0198* |
| **LARO (TFF=1, SPO=2)** | **43.38 ± 2.94** | **0.9751 ± 0.0142** |

| R2* | PSNR (↑) | SSIM (↑) |
|---|---|---|
| TFF=0, SPO=0 | 38.83 ± 3.19* | 0.9447 ± 0.0297* |
| TFF=0, SPO=1 | 39.69 ± 3.32* | 0.9463 ± 0.0289* |
| TFF=1, SPO=0 | 39.35 ± 3.17* | 0.9513 ± 0.0267* |
| TFF=1, SPO=1 | 40.35 ± 3.35* | 0.9539 ± 0.0259* |
| **LARO (TFF=1, SPO=2)** | **41.59 ± 3.41** | **0.9644 ± 0.0197** |

| RDF | PSNR (↑) | SSIM (↑) |
|---|---|---|
| TFF=0, SPO=0 | 32.98 ± 3.25* | 0.8675 ± 0.0522* |
| TFF=0, SPO=1 | 33.76 ± 3.23* | 0.8710 ± 0.0526* |
| TFF=1, SPO=0 | 33.36 ± 3.16* | 0.8805 ± 0.0471* |
| TFF=1, SPO=1 | 34.21 ± 3.14* | 0.8844 ± 0.0469* |
| **LARO (TFF=1, SPO=2)** | **35.21 ± 3.38** | **0.9058 ± 0.0376** |

| QSM | RMSE (↓) | SSIM (↑) | HFEN (↓) |
|---|---|---|---|
| TFF=0, SPO=0 | 49.48 ± 4.39* | 0.9735 ± 0.0088 | 36.13 ± 1.14* |
| TFF=0, SPO=1 | 46.27 ± 3.43* | 0.9778 ± 0.0016* | 34.97 ± 2.05* |
| TFF=1, SPO=0 | 47.57 ± 4.34* | 0.9813 ± 0.0026 | 33.92 ± 1.79* |
| TFF=1, SPO=1 | 44.11 ± 3.49 | 0.9802 ± 0.0029 | 33.61 ± 2.09 |
| **LARO (TFF=1, SPO=2)** | **42.51 ± 4.12** | **0.9881 ± 0.0022** | **31.60 ± 2.14** |

Table S2. Performance comparison on retrospectively under-sampled acquired k-space data (800 2D coronal slices of magnitude, R2* and RDF, and 4 3D volumes of QSM) with acceleration factor $R = 8$. For MoDL and TFF, reconstruction accuracies (mean ± standard) of four maps were progressively improved from sampling pattern SPO=0, 1 to 2. Under each sampling pattern, TFF consistently outperformed MoDL and LLR (*: statistical significance compared to TFF under each sampling pattern, p < 0.05).

| Magnitude | | PSNR (↑) | SSIM (↑) |
|---|---|---|---|
| SPO=0 | LLR | 33.76 ± 3.39* | 0.9132 ± 0.0492* |
| | MoDL | 39.30 ± 3.02* | 0.9550 ± 0.0250* |
| | **TFF** | **40.13 ± 3.14** | **0.9616 ± 0.0211** |
| SPO=1 | LLR | 34.00 ± 3.74* | 0.9104 ± 0.0513* |
| | MoDL | 40.77 ± 2.90* | 0.9598 ± 0.0233* |
| | **TFF** | **41.65 ± 2.89** | **0.9656 ± 0.0198** |
| SPO=2 | LLR | 34.67 ± 3.36* | 0.9268 ± 0.0424* |
| | MoDL | 42.60 ± 2.74* | 0.9717 ± 0.0157* |
| | **LARO** | **43.38 ± 2.94** | **0.9751 ± 0.0142** |

| R2* | | PSNR (↑) | SSIM (↑) |
|---|---|---|---|
| SPO=0 | LLR | 28.35 ± 6.87* | 0.8858 ± 0.0648* |
| | MoDL | 38.51 ± 3.17* | 0.9442 ± 0.0311* |
| | **TFF** | **39.35 ± 3.17** | **0.9513 ± 0.0267** |
| SPO=1 | LLR | 28.13 ± 6.79* | 0.8811 ± 0.0672* |
| | MoDL | 39.35 ± 3.24* | 0.9457 ± 0.0313* |
| | **TFF** | **40.35 ± 3.35** | **0.9539 ± 0.0259** |
| SPO=2 | LLR | 28.31 ± 6.92* | 0.8909 ± 0.0639* |
| | MoDL | 40.99 ± 3.45* | 0.9614 ± 0.0207* |
| | **LARO** | **41.59 ± 3.41** | **0.9644 ± 0.0197** |

| RDF | | PSNR (↑) | SSIM (↑) |
|---|---|---|---|
| SPO=0 | LLR | 30.12 ± 4.43* | 0.7866 ± 0.1163* |
| | MoDL | 32.86 ± 3.19* | 0.8704 ± 0.0518* |
| | **TFF** | **33.36 ± 3.16** | **0.8805 ± 0.0471** |
| SPO=1 | LLR | 30.05 ± 4.59* | 0.7803 ± 0.1167* |
| | MoDL | 33.67 ± 3.21* | 0.8757 ± 0.0515* |
| | **TFF** | **34.21 ± 3.14** | **0.8844 ± 0.0469** |
| SPO=2 | LLR | 30.16 ± 5.11* | 0.7907 ± 0.1179* |
| | MoDL | 34.74 ± 3.49* | 0.9006 ± 0.0392* |
| | **LARO** | **35.21 ± 3.38** | **0.9058 ± 0.0376** |

| QSM | | RMSE (↓) | SSIM (↑) | HFEN (↓) |
|---|---|---|---|---|
| SPO=0 | LLR | 58.26 ± 6.29 | 0.9583 ± 0.0145 | 49.32 ± 4.26 |
| | MoDL | 62.66 ± 12.25 | 0.9774 ± 0.0045 | 52.41 ± 13.50 |
| | **TFF** | **47.57 ± 4.34** | **0.9813 ± 0.0026** | **33.92 ± 1.79** |
| SPO=1 | LLR | 58.78 ± 5.46 | 0.9573 ± 0.0125 | 53.28 ± 4.59 |
| | MoDL | 46.01 ± 2.62 | 0.9744 ± 0.0059 | 35.88 ± 2.59 |
| | **TFF** | **44.11 ± 3.49** | **0.9802 ± 0.0029** | **33.61 ± 2.09** |
| SPO=2 | LLR | 57.76 ± 7.12 | 0.9659 ± 0.0054 | 48.90 ± 5.45 |
| | MoDL | 44.22 ± 5.40 | 0.9868 ± 0.0037 | 32.95 ± 1.84 |

| | LARO | 42.51 ± 4.12 | 0.9881 ± 0.0022 | 31.60 ± 2.14 |

Table S3. Ablation study on retrospectively under-sampled synthetic k-space data (800 2D coronal slices of magnitude, R2* and RDF, and 4 3D volumes of QSM) with acceleration factor $R = 4$. Reconstruction accuracies (mean ± standard) of all maps were progressively improved as more modules were added, where LARO (TFF=1, SPO=2) performed the best (*: statistical significance compared to LARO, $p < 0.05$).

| Magnitude | PSNR (↑) | SSIM (↑) |
| --- | --- | --- |
| TFF=0, SPO=0 | 37.83 ± 3.32* | 0.9419 ± 0.0327* |
| TFF=0, SPO=1 | 38.66 ± 3.44* | 0.9505 ± 0.0281* |
| TFF=1, SPO=0 | 40.02 ± 3.10* | 0.9661 ± 0.0195* |
| TFF=1, SPO=1 | 40.42 ± 3.02* | 0.9694 ± 0.0175* |
| **LARO (TFF=1, SPO=2)** | **41.51 ± 2.95** | **0.9754 ± 0.0135** |

| R2* | PSNR (↑) | SSIM (↑) |
| --- | --- | --- |
| TFF=0, SPO=0 | 38.03 ± 2.99* | 0.9468 ± 0.0298* |
| TFF=0, SPO=1 | 38.80 ± 3.02* | 0.9539 ± 0.0260* |
| TFF=1, SPO=0 | 39.86 ± 3.01* | 0.9632 ± 0.0216* |
| TFF=1, SPO=1 | 40.26 ± 2.90* | 0.9665 ± 0.0193* |
| **LARO (TFF=1, SPO=2)** | **41.20 ± 2.99** | **0.9723 ± 0.0155** |

| RDF | PSNR (↑) | SSIM (↑) |
| --- | --- | --- |
| TFF=0, SPO=0 | 33.14 ± 3.38* | 0.8715 ± 0.05835* |
| TFF=0, SPO=1 | 33.84 ± 3.24* | 0.8889 ± 0.050* |
| TFF=1, SPO=0 | 35.18 ± 3.30* | 0.9069 ± 0.0420* |
| TFF=1, SPO=1 | 35.50 ± 3.24* | 0.9156 ± 0.0367* |
| **LARO (TFF=1, SPO=2)** | **36.39 ± 3.58** | **0.9259 ± 0.0313** |

| QSM | RMSE (↓) | SSIM (↑) | HFEN (↓) |
|---|---|---|---|
| TFF=0, SPO=0 | 40.30 ± 3.80* | 0.9731 ± 0.0108* | 29.48 ± 1.85* |
| TFF=0, SPO=1 | 37.41 ± 2.40* | 0.9769 ± 0.0100* | 25.50 ± 1.72* |
| TFF=1, SPO=0 | 33.57 ± 1.46 | 0.9860 ± 0.0056 | 22.22 ± 2.78 |
| TFF=1, SPO=1 | 32.62 ± 2.23 | 0.9871 ± 0.0048 | 20.23 ± 1.34 |
| **LARO (TFF=1, SPO=2)** | **30.77 ± 3.20** | **0.9890 ± 0.0036** | **19.06 ± 1.45** |

Table S4. Performance comparison on retrospectively under-sampled synthetic k-space data (800 2D coronal slices of magnitude, R2* and RDF, and 4 3D volumes of QSM) with acceleration factor $R = 4$. For each method, reconstruction accuracies (mean ± standard) of all maps were progressively improved from sampling pattern SPO=0, 1 to 2. Under each sampling pattern, TFF consistently outperformed MoDL and LLR (*: statistical significance compared to TFF under each sampling pattern, p < 0.05).

| Magnitude | | PSNR (↑) | SSIM (↑) |
|---|---|---|---|
| SPO=0 | LLR | 32.85 ± 3.67* | 0.9081 ± 0.0541* |
| | MoDL | 38.29 ± 3.26* | 0.9544 ± 0.0287* |
| | **TFF** | **40.21 ± 3.14** | **0.9668 ± 0.0191** |
| SPO=1 | LLR | 33.98 ± 3.75* | 0.9218 ± 0.0471* |
| | MoDL | 39.01 ± 3.34* | 0.9604 ± 0.0250* |
| | **TFF** | **40.34 ± 3.17** | **0.9693 ± 0.0174** |
| SPO=2 | LLR | 34.97 ± 3.71* | 0.9342 ± 0.0409* |
| | MoDL | 39.94 ± 3.21* | 0.9682 ± 0.0182* |
| | **LARO** | **41.53 ± 3.02** | **0.9756 ± 0.0136** |

| R2* | | PSNR (↑) | SSIM (↑) |
|---|---|---|---|
| SPO=0 | LLR | 34.95 ± 3.06* | 0.9203 ± 0.0422* |
| | MoDL | 38.55 ± 3.18* | 0.9551 ± 0.0270* |
| | **TFF** | **39.94 ± 2.94** | **0.9636 ± 0.0210** |
| SPO=1 | LLR | 36.06 ± 3.06* | 0.9322 ± 0.0359* |
| | MoDL | 39.14 ± 3.03* | 0.9603 ± 0.0241* |
| | **TFF** | **40.21 ± 2.96** | **0.9666 ± 0.0188** |
| SPO=2 | LLR  | 37.03 ± 3.22*     | 0.9430 ± 0.0306*    |
|       | MoDL | 40.04 ± 3.04*     | 0.9676 ± 0.0183*    |
|       | **LARO** | **41.27 ± 2.97** | **0.9728 ± 0.0152** |

| RDF |      | PSNR (↑)          | SSIM (↑)            |
|-----|------|-------------------|---------------------|
| SPO=0 | LLR  | 30.84 ± 3.55*   | 0.8226 ± 0.0762*    |
|       | MoDL | 33.95 ± 3.70*   | 0.8944 ± 0.0481*    |
|       | **TFF** | **35.18 ± 3.46** | **0.9065 ± 0.0411** |
| SPO=1 | LLR  | 31.88 ± 3.58*   | 0.8520 ± 0.0624*    |
|       | MoDL | 34.43 ± 3.56*   | 0.9044 ± 0.0430*    |
|       | **TFF** | **35.35 ± 3.37** | **0.9137 ± 0.0371** |
| SPO=2 | LLR  | 32.80 ± 3.49*   | 0.8707 ± 0.0548*    |
|       | MoDL | 35.24 ± 3.68*   | 0.9169 ± 0.0353*    |
|       | **LARO** | **36.36 ± 3.67** | **0.9258 ± 0.0314** |

| QSM |      | RMSE (↓)          | SSIM (↑)            | HFEN (↓)            |
|-----|------|-------------------|---------------------|---------------------|
| SPO=0 | LLR  | 49.44 ± 5.86*   | 0.9570 ± 0.0208*    | 41.92 ± 1.50*       |
|       | MoDL | 35.85 ± 1.07    | 0.9807 ± 0.0071     | 24.46 ± 2.46*       |
|       | **TFF** | **32.89 ± 2.15** | **0.9826 ± 0.0066** | **21.47 ± 1.37** |
| SPO=1 | LLR  | 44.89 ± 7.05*   | 0.9671 ± 0.0141*    | 33.51 ± 1.31*       |
|       | MoDL | 33.47 ± 0.92    | 0.9826 ± 0.0069     | 21.90 ± 1.87        |
|       | **TFF** | **32.08 ± 1.64** | **0.9838 ± 0.0060** | **19.73 ± 0.74** |
| SPO=2 | LLR  | 39.47 ± 4.45*   | 0.9731 ± 0.0121*    | 28.17 ± 1.89*       |
|       | MoDL | 31.69 ± 1.51    | 0.9858 ± 0.0061     | 19.60 ± 0.93        |

| | LARO | 29.34 ± 2.15 | 0.9874 ± 0.0044 | 17.72 ± 1.02 |
|---|---|---|---|---|

Table S5. Mean susceptibility values and standard deviations in Globus pallidus (GP), Substantia Nigra (SN), Red Nucleus (RN), Caudate Nucleus (CN), Putamen (PU), thalamus (TH), optic radiation (OR) and cerebral cortex (CC) for R=2 SENSE and R=8 accelerated reconstructions. Compared to SENSE as reference, some under-estimations in SN, RN, CN and CC reconstructed by MoDL and TFF were seen when SPO=0 and 1 but were recovered when SPO=2. Compared to MoDL and TFF, LLR had more deviations from SENSE.

|  |  | GP | SN | RN | CN | PU | TH | OR | CC |
|---|---|---|---|---|---|---|---|---|---|
| **R=2, SENSE** |  | 0.153 ± 0.029 | 0.143 ± 0.041 | 0.091 ± 0.019 | 0.053 ± 0.016 | 0.034 ± 0.016 | 0.000 ± 0.014 | -0.058 ± 0.020 | -0.010 ± 0.015 |
| **R=8, SPO=0** | LLR | 0.130 ± 0.041 | 0.106 ± 0.059 | 0.061 ± 0.029 | 0.047 ± 0.024 | 0.035 ± 0.026 | -0.002 ± 0.017 | -0.046 ± 0.016 | -0.040 ± 0.031 |
|  | MoDL | 0.143 ± 0.034 | 0.124 ± 0.044 | 0.086 ± 0.028 | 0.041 ± 0.016 | 0.036 ± 0.018 | 0.001 ± 0.015 | -0.053 ± 0.016 | -0.033 ± 0.028 |
|  | TFF | 0.147 ± 0.033 | 0.123 ± 0.049 | 0.083 ± 0.027 | 0.042 ± 0.015 | 0.039 ± 0.017 | 0.001 ± 0.014 | -0.050 ± 0.016 | -0.029 ± 0.024 |
| **R=8, SPO=1** | LLR | 0.129 ± 0.045 | 0.101 ± 0.058 | 0.063 ± 0.025 | 0.045 ± 0.024 | 0.039 ± 0.025 | 0.003 ± 0.018 | -0.041 ± 0.016 | -0.040 ± 0.034 |
|  | MoDL | 0.142 ± 0.033 | 0.122 ± 0.048 | 0.078 ± 0.033 | 0.047 ± 0.017 | 0.035 ± 0.017 | -0.001 ± 0.016 | -0.056 ± 0.017 | -0.044 ± 0.024 |
|  | TFF | 0.147 ± 0.030 | 0.121 ± 0.049 | 0.079 ± 0.025 | 0.044 ± 0.016 | 0.038 ± 0.017 | -0.003 ± 0.014 | -0.059 ± 0.016 | -0.027 ± 0.020 |
| **R=8, SPO=2** | LLR | 0.126 ± 0.049 | 0.112 ± 0.059 | 0.066 ± 0.023 | 0.051 ± 0.038 | 0.043 ± 0.034 | -0.002 ± 0.023 | -0.031 ± 0.021 | -0.036 ± 0.032 |
|  | MoDL | 0.145 ± 0.034 | 0.133 ± 0.050 | 0.094 ± 0.024 | 0.046 ± 0.016 | 0.036 ± 0.016 | -0.001 ± 0.015 | -0.056 ± 0.017 | -0.023 ± 0.020 |
|  | LARO | 0.150 ± 0.030 | 0.141 ± 0.051 | 0.092 ± 0.025 | 0.052 ± 0.015 | 0.037 ± 0.017 | 0.001 ± 0.015 | -0.053 ± 0.018 | -0.008 ± 0.015 |